\documentclass[12pt]{article}
\usepackage{latexsym}
\usepackage{amsmath,amsfonts}
\usepackage{times}
\allowdisplaybreaks[4]

\catcode`\@=11

\newcount\hour
\newcount\minute
\newtoks\amorpm \hour=\time\divide\hour by 60\minute
=\time{\multiply\hour by 60 \global\advance\minute by-\hour}
\edef\standardtime{{\ifnum\hour<12 \global\amorpm={am}%
        \else\global\amorpm={pm}\advance\hour by-12 \fi
        \ifnum\hour=0 \hour=12 \fi
        \number\hour:\ifnum\minute<10
        0\fi\number\minute\the\amorpm}}
\edef\militarytime{\number\hour:\ifnum\minute<10 0\fi\number\minute}

\def\draftlabel#1{{\@bsphack\if@filesw {\let\thepage\relax
   \xdef\@gtempa{\write\@auxout{\string
      \newlabel{#1}{{\@currentlabel}{\thepage}}}}}\@gtempa
   \if@nobreak \ifvmode\nobreak\fi\fi\fi\@esphack}
        \gdef\@eqnlabel{#1}}
\def\@eqnlabel{}
\def\@vacuum{}
\def\marginnote#1{}
\def\draftmarginnote#1{\marginpar{\raggedright\scriptsize\tt#1}}
\def\draft{
        \pagestyle{plain}
        \overfullrule=2pt
        \oddsidemargin -.5truein
        \def\@oddhead{\sl \phantom{\today\quad\militarytime} \hfil
        \smash{\Large\sl DRAFT} \hfil \today\quad\militarytime}
        \let\@evenhead\@oddhead
        \let\label=\draftlabel
        \let\marginnote=\draftmarginnote
        \def\ps@empty{\let\@mkboth\@gobbletwo
        \def\@oddfoot{\hfil \smash{\Large\sl DRAFT} \hfil}
        \let\@evenfoot\@oddhead}
        \def\@eqnnum{(\theequation)\rlap{\kern\marginparsep\tt\@eqnlabel}%
        \global\let\@eqnlabel\@vacuum}  }

\newcommand{\rf}[1]{(\ref{#1})}
\renewcommand{\theequation}{\thesection.\arabic{equation}}
\renewcommand{\thefootnote}{\fnsymbol{footnote}}
\newcommand{\newsection}{   
\setcounter{equation}{0}\section}

\def\appendix#1{\addtocounter{section}{1}\setcounter{equation}{0}
\renewcommand{\thesection}{\Alph{section}}
\section*{Appendix \thesection\protect\indent \parbox[t]{11.15cm}{#1}}
\addcontentsline{toc}{section}{Appendix \thesection\ \ \ #1}}

\def\be{\begin{equation}}
\def\ee{\end{equation}}
\def\beq{\begin{eqnarray}}
\def\eeq{\end{eqnarray}}

\def\parline{\,\partial\kern -0.55em /\,\,}

\def\half{{\frac{1}{2}}}

\def\CC{{\cal C}}

\def\LL{{\cal L}}
\def\MM{{\cal M}}

\def\TT{{\cal T}}

\def\phik{|\phi\rangle}
\def\phibr{\langle\phi|}

\def\smponetwo{{\scriptscriptstyle [1,2]}}

\def\oplussm{{\scriptscriptstyle \oplus}}
\def\ominussm{{\scriptscriptstyle \ominus}}

\def\oplussm{{\scriptscriptstyle \oplus}}
\def\ominussm{{\scriptscriptstyle \ominus}}

\def\m{{\rm m}}

\def\eb{{\bar{e}}}

\def\rb{{\bar{r}}}

\def\alphabf{{\boldsymbol{\alpha}}}
\def\phibf{{\boldsymbol{\phi}}}
\def\xibf{{\boldsymbol{\xi}}}
\def\mubf{{\boldsymbol{\mu}}}
\def\Pibf{{\boldsymbol{\Pi}}}
\def\partialbf{{\boldsymbol{\partial}}}

\def\Abf{{\bf A}}

\def\Lbf{{\bf L}}

\newcommand{\mc}{\multicolumn}

\def\eff{{\rm eff}}

\def\sh{{\rm sh}}

\def\cur{{\rm cur}}
\def\cur{{\rm cur}}
\def\diff{{\rm diff}}
\def\field{{\rm field}}

\begin{document}


\begin{flushright}
FIAN-TD-2013-19 \hspace{1.8cm}{}~\\
arXiv: 1312.5679 V2 [hep-th]  {}~ \\
\end{flushright}

\vspace{1cm}

\begin{center}

{\Large \bf Light-cone gauge approach to arbitrary spin fields, currents, and shadows

\bigskip  }

\vspace{2.5cm}

R.R. Metsaev\footnote{ E-mail: metsaev@lpi.ru }

\vspace{1cm}

{\it Department of Theoretical Physics, P.N. Lebedev Physical Institute, \\
Leninsky prospect 53,  Moscow 119991, Russia }

\vspace{3.5cm}

{\bf Abstract}

\end{center}

Totally symmetric arbitrary spin fields in
AdS space, conformal fields, conformal currents, and shadow fields in flat space are studied.
Light-cone gauge formulation for such fields, currents, and shadows is obtained. Use of the Poincar\'e parametrization of AdS space and ladder operators allows us to treat fields in flat and AdS spaces on equal footing. Light-cone gauge realization of relativistic symmetries for fields, currents, and shadows is also obtained. The light-cone gauge formulation for fields is obtained by using the gauge invariant Lagrangian which is presented in terms of modified de Donder divergence, while the light-cone gauge formulation for currents and shadows is obtained by using the gauge invariant approach to currents and shadows. This allows us to  demonstrate explicitly how the ladder operators entering the gauge invariant formulation of fields, currents, and shadows manifest themselves in the light-cone gauge formulation for fields, currents, and shadows.

\newpage
\renewcommand{\thefootnote}{\arabic{footnote}}
\setcounter{footnote}{0}

\section{\large Introduction}

The light-cone gauge formalism \cite{Dirac:1951zz} provides systematic and self-contained way to study many problems of field and string theories. For example, we mention the construction of superfield formulation for some supersymmetric theories \cite{Mandelstam:1982cb} and  light-cone gauge string field theory \cite{kaku,Green:1983hw}. Another application of the light-cone gauge formalism is a construction of interaction vertices in the theory of higher-spin fields
\cite{Bengtsson:1983pd}-\cite{Metsaev:2005ar}. Some attractive applications of the light-cone gauge formalism to theories like QCD may be found in Refs.\cite{Brodsky:1997de}.

In this paper, we develop light-cone gauge approach to fields in AdS space, conformal fields, conformal currents, and shadow fields in flat space.%
\footnote{ For the first time, light-cone gauge approach to AdS fields, currents, and shadows was developed in Refs.\cite{Metsaev:1999ui,Metsaev:2005ws}. Application of light-cone approach in Ref.\cite{Metsaev:1999ui} to the various AdS field dynamical systems and study of AdS/CFT correspondence may be found in Refs.\cite{Metsaev:1999gz}-\cite{Metsaev:2004ee}. Other interesting application of formalism in Ref.\cite{Metsaev:1999ui} to the study of AdS/CFT correspondence may be found in Refs.\cite{Koch:2010cy}. Advantage of light-cone approach to AdS field dynamics we develop in this paper as compared to the one in Ref.\cite{Metsaev:1999ui} is that light-cone gauge action obtained in this paper leads to decoupled equations of motion which are easily solved in terms of Bessel functions. Advantage of light-cone approach to currents and shadows in this paper as compared the one in Ref.\cite{Metsaev:2005ws} is that light-cone gauge generators of conformal algebra obtained in this paper are polynomial with respect to derivatives  $\partial^i$, $\partial^-$. Light-cone gauge formulation of conformal fields developed in this paper has not been discussed in earlier literature.}
Our approach to the light-cone gauge formulation of fields, currents, and shadows can be summarized as follows.

\noindent {\bf i}) To obtain light-cone gauge formalism for the fields, we use Lagrangian gauge invariant formulation of AdS fields developed in Refs.\cite{Metsaev:2008ks,Metsaev:2009hp} and gauge invariant Lagrangian ordinary-derivative formulation of conformal fields in flat space developed in Refs.\cite{Metsaev:2007fq,Metsaev:2007rw}. Our representation for gauge invariant Lagrangian is based on the use of ladder operators and modified de Donder divergence.
The massless and massive fields in $AdS_{d+1}$ space are considered by using the Poincar\'e parametrization of $AdS_{d+1}$, while the
conformal fields in flat space $R^{d-1,1}$ are considered by using the Cartesian parametrization of $R^{d-1,1}$. In our gauge invariant approach, we use the double-traceless tensor fields of the Lorentz algebra $so(d-1,1)$ and the $so(d-1,1)$ symmetries are manifestly realized both for AdS fields and conformal fields, while, in our light-cone gauge approach, we use traceless tensor fields of the $so(d-2)$ algebra and the $so(d-2)$ symmetries are manifestly realized both for light-cone gauge AdS fields and conformal fields. We note that it is the use of the ladder operators that allows us to develop the  light-cone gauge formulation of fields in AdS space and conformal fields in flat space on equal footing.

\noindent {\bf ii}) To develop light-cone gauge formalism for the currents and shadows, we use gauge invariant approach to currents and shadows developed in Refs.\cite{Metsaev:2008fs}-\cite{Metsaev:2011uy}. This gauge invariant approach turns out to be convenient for the derivation of  light-cone gauge approach to currents and shadows.

Our paper  is organized as follows.

In Sec.\ref{section-02}, we review the Lagrangian gauge invariant formulation of massless and massive fields in flat and AdS spaces and conformal
fields in flat space. Representation for gauge invariant Lagrangian in
terms of the modified de Donder divergence discovered in
Refs.\cite{Metsaev:2008ks}-\cite{Metsaev:2007rw} is discussed. In Sec.\ref{section-03}, we review  the gauge invariant approach to currents and shadows. In Sec.\ref{section-04}, we describe  $so(d-1,1)$ covariant realization of relativistic symmetries of fields, currents, and shadows. In Sec.\ref{section-05}, we develop the light-cone gauge formulation of fields, currents, and shadows. First, we describe the field contents
appearing in our light-cone gauge formulation. After this, we present our result for
light-cone gauge action for fields and 2-point vertices for currents and shadows.  In Sec.\ref{section-06}, we discuss the light-cone gauge realization of relativistic symmetries for fields, currents, and shadows. In Sec.\ref{section-07}, we briefly discuss some potentially interesting applications of our results. In Appendix, we summarize our conventions and the notation.

\newsection{\large Gauge invariant approach to fields }
\label{section-02}

We obtain our light-cone gauge formulation by using Lagrangian gauge invariant approach. We start therefore with review  of gauge invariant metric-like Lagrangian approach to totally symmetric fields in flat and AdS spaces.%
\footnote{ For massless and massive fields in flat space, derivation of light-cone gauge approach from Lagrangian BRST approach may be found in Refs.\cite{Siegel:1999ew}.}
We discuss the dynamics of the following totally symmetric  fields:

\noindent {\bf i}) massless and massive spin-$s$ fields in $R^{d-1,1}$;

\noindent {\bf ii}) massless and massive spin-$s$ fields in $AdS_{d+1}$;

\noindent {\bf iii}) conformal spin-$s$ field in $R^{d-1,1}$.

We use the following parametrizations of flat space $R^{d-1,1}$ and $AdS_{d+1}$ space,
\beq
\label{man02-13122013-01} && ds^2  =  dx^a dx^a\,, \hspace{5.1cm} \hbox{ for flat space},
\\
\label{man02-13122013-02} && ds^2 = \frac{1}{z^2} (dx^a dx^a + dzdz)\,,   \hspace{3cm} \hbox{ for AdS
space}.
\eeq
The manifest symmetries of line elements in \rf{man02-13122013-01},\rf{man02-13122013-02} are described by the Lorentz algebra $so(d-1,1)$. It is the use of the manifest $so(d-1,1)$ symmetries that allows us, among other things, to treat fields in flat and AdS spaces on equal footing. We now discuss field contents.

{\bf Field contents}. To discuss Lagrangian description of above-mentioned fields we use totally symmetric double-traceless tensor fields  of the $so(d-1,1)$ algebra. The field contents we use are presented in Table I. To simplify the presentation of gauge invariant action we use oscillators $\alpha^a$, $\alpha^z$, $\zeta$, $\upsilon^\oplussm$, $\upsilon^\ominussm$  and introduce the corresponding ket-vector which are also presented in Table I.

Concerning the field contents in Table I, the following remarks are in order.

\noindent{\bf i}) Lagrangian description of massless spin-$s$ field
in $R^{d-1,1}$ with the field content given in Table I was developed in Ref.\cite{Fronsdal:1978rb}.

\noindent{\bf ii}) Lagrangian description of massive spin-$s$ field
in $R^{d-1,1}$ with the field content given in Table I was discussed in Ref.\cite{Zinoviev:2001dt}. Below we discuss presentation of Lagrangian for massive field in terms of de Donder divergence which was obtained in Ref.\cite{Metsaev:2008fs}.%
\footnote{ In earlier literature, the derivation of Lagrangian for massive field by using the dimensional reduction may be found in Refs.\cite{Aragone:1988yx,Rindani:1988gb}. BRST approach to massive field in flat space is studied in Ref.\cite{Buchbinder:2005ua}.}

\noindent{\bf iii}) Lagrangian description of massless spin-$s$ field
in $AdS_{d+1}$ with the field content given in Table I was discussed in Ref.\cite{Metsaev:2008ks}. For the first time, Lagrangian description of massless field in $AdS_{d+1}$, $d=3$, was obtained in Ref.\cite{Fronsdal:1978vb} by using totally symmetric double-traceless tensor field of the Lorentz algebra $so(d,1)$. The $so(d-1,1)$ tensorial components of the $so(d,1)$ tensor field in Ref.\cite{Fronsdal:1978vb} are not double-traceless and this tensor field is related to our gauge fields given in Table I by invertible transformation described in Ref.\cite{Metsaev:2008ks}.%
\footnote{ Frame-like approach to massless AdS fields was discussed in Refs.\cite{Lopatin:1987hz}. For arbitrary $d$, metric-like gauge invariant formulation of
massless AdS fields was discussed in Refs.\cite{Metsaev:1999ui,Buchbinder:2001bs}.
Discussion  of various Lagrangian formulations of higher-spin field dynamics
in terms of unconstrained fields may be found in Refs.\cite{Sagnotti:2003qa}-\cite{Buchbinder:2008ss}. }

\noindent{\bf iv}) Lagrangian description of massive spin-$s$ field
in $AdS_{d+1}$ with the field content given in Table I was discussed in Ref.\cite{Metsaev:2009hp}. In Ref.\cite{Zinoviev:2001dt}, massive field
in $AdS_{d+1}$ is described by the set of fields involving
double-traceless tensor fields of the Lorentz algebra
$so(d,1)$. The $so(d-1,1)$ tensorial components of the tensor
fields in Ref.\cite{Zinoviev:2001dt} are not double-traceless. The
fields in Ref.\cite{Zinoviev:2001dt} are related to our fields
given in Table I by invertible transformation
described in Ref.\cite{Metsaev:2009hp}.%
\footnote{ Frame-like approach to massive AdS fields was discussed in Refs.\cite{Zinoviev:2008ze,Ponomarev:2010st}. BRST approach to massive fields is studied in Refs.\cite{Buchbinder:2006ge}-\cite{Grigoriev:2011gp}. Discussion of massive AdS fields
via various dimensional reduction techniques may be found in
Refs.\cite{Metsaev:2000qb}-\cite{Artsukevich:2008vy}.}

{\small
\noindent {\sf Table I. Field contents and the corresponding ket-vectors entering gauge invariant formulation of fields, currents, and shadows. Fields with $s'\geq 4$ are totally symmetric double-traceless tensor fields of $so(d-1,1)$ algebra, $\phi^{aabba_5\ldots a_{s'}}=0$, $s'\geq 4$, i.e., $(\bar\alphabf^2)^2|\phibf\rangle = 0$.  For notion of $\lambda \in [s-s']_2$ and oscillator algebra see Appendix. In the Table, algebraic constraints express the homogeneity properties of ket-vectors $|\phibf\rangle$ with respect to the oscillators.}}
\begin{center}
\begin{tabular}{|l|c|l|}
\hline &&
\\[-3mm]
 & Field content   & Ket-vector $|\phibf\rangle$ and algebraic constraints
\\ [1mm]\hline
&&
\\[-1mm]
massless spin-$s$ & $\phibf^{a_1\ldots a_s}$  & $\frac{1}{s!} \alpha^{a_1} \ldots \alpha^{a_s}
\phibf^{a_1\ldots a_s} |0\rangle$, \hspace{1.5cm} $(N_\alphabf-s)|\phibf\rangle=0$
\\[1mm]
field in $R^{d-1,1}$ &&
\\ [1mm]\hline
&&
\\[-3mm]
massive spin-$s$ & $\phibf^{a_1\ldots a_{s'}}$  & $
\sum\limits_{s'=0}^s \frac{\zeta^{s-s'} \alpha^{a_1} \ldots
\alpha^{a_{s'}}}{s'!\sqrt{(s - s')!}} \, \phibf^{a_1\ldots a_{s'}} |0\rangle$,\qquad  $(N_\alphabf+N_\zeta-s)|\phibf\rangle=0$
\\[2mm]
field in $R^{d-1,1}$ & {\small $ s'=0,1,\ldots, s $} &
\\ [1mm]\hline
&&
\\[-3mm]
massless spin-$s$ & $\phibf^{a_1\ldots a_{s'}}$  & $
\sum\limits_{s'=0}^s \frac{\alpha_z^{s-s'} \alpha^{a_1} \ldots
\alpha^{a_{s'}}}{s'!\sqrt{(s - s')!}} \, \phibf^{a_1\ldots a_{s'}} |0\rangle$, \hspace{0.5cm} $(N_\alphabf + N_z-s)|\phibf\rangle=0$
\\[2mm]
field in $AdS_{d+1} $ & {\small $ s'=0,1,\ldots, s $} &
\\ [1mm]\hline
&&
\\[-3mm]
massive spin-$s$ & $\phibf_\lambda^{a_1\ldots a_{s'}}$  & $
\sum\limits_{s'=0}^s\,\,\sum\limits_{\lambda\in [s-s']_2}
\frac{\zeta_{\phantom{z}}^{\frac{s-s'+ \lambda}{2}}
\alpha_z^{\frac{s-s'- \lambda}{2}}\alpha^{a_1}\ldots
\alpha^{a_{s'}}}{s'!\sqrt{(\frac{s-s'+ \lambda}{2})! (\frac{s-s'- \lambda}{2})!}}
\, \phibf_\lambda^{a_1\ldots a_{s'}} |0\rangle$
\\[2mm]
field in $AdS_{d+1} $ & {\small $ s'=0,1,\ldots, s $} &
\\ [1mm]
& {\small $\lambda \in [s-s']_2$ }&  \hspace{2cm} $(N_\alphabf + N_z+ N_\zeta-s)|\phibf\rangle=0$
\\ [1mm]\hline
&&
\\[-3mm]
conformal spin-$s$ & $\phibf_{k'}^{a_1\ldots a_{s'}}$  & $
\sum\limits_{s'=0}^s \sum\limits_{k'\in
[k_{s'}]_2} \!\!\! \frac{\zeta^{s-s'}
(\upsilon^\ominussm)^{^{\frac{k_{s'}+k'}{2}}}
(\upsilon^\oplussm)^{^{\frac{k_{s'} - k'}{2}}}}{s'!
\sqrt{(s-s')!}(\frac{k_{s'} + k'}{2})!}\alpha^{a_1} \ldots
\alpha^{a_{s'}} \, \phibf_{k'}^{a_1\ldots a_{s'}}|0\rangle$
\\[2mm]
field in $R^{d-1,1}$ &  {\small $ s'=0,1,\ldots, s $}  &
\\ [1mm]
& {\small $ \hspace{-0.8cm}  k' \in [k_{s'}]_2$ }  &  \hspace{0.3cm} $(N_\alphabf + N_\zeta-s)|\phibf\rangle=0$,  \hspace{0.5cm} $(N_\zeta + N_\upsilon  - k_s)|\phibf\rangle=0$
\\ [1mm]
& {\small $k_{s'} \equiv s' + \frac{d-6}{2}$ } &
\\ [1mm]\hline
&&
\\[-3mm]
canonical spin-$s$ & $\phibf_\cur^{a_1\ldots a_{s'}}$  & $
\sum\limits_{s'=0}^s \frac{\alpha_z^{s-s'} \alpha^{a_1} \ldots
\alpha^{a_{s'}}}{s'!\sqrt{(s - s')!}} \, \phibf_\cur^{a_1\ldots a_{s'}} |0\rangle$,  \hspace{0.5cm} $(N_\alphabf + N_z - s)|\phibf\rangle=0$
\\[2mm]
current in $R^{d-1,1}$& {\small $ s'=0,1,\ldots, s $} &
\\ [1mm]\hline
&&
\\[-3mm]
canonical spin-$s$ & $\phibf_\sh^{a_1\ldots a_{s'}}$  & $
\sum\limits_{s'=0}^s \frac{\alpha_z^{s-s'} \alpha^{a_1} \ldots
\alpha^{a_{s'}}}{s'!\sqrt{(s - s')!}} \, \phibf_\sh^{a_1\ldots a_{s'}} |0\rangle$,  \hspace{0.5cm} $(N_\alphabf + N_z - s)|\phibf\rangle=0$
\\[2mm]
shadow in $R^{d-1,1}$ & {\small $ s'=0,1,\ldots, s $} &
\\ [1mm]\hline
&&
\\[-3mm]
anomalous spin-$s$ & $\phibf_{\cur, \lambda}^{a_1\ldots a_{s'}}$  & $
\sum\limits_{s'=0}^s\,\,\sum\limits_{\lambda \in [s-s']_2}
\frac{\zeta_{\phantom{z}}^{\frac{s-s'+\lambda}{2}}
\alpha_z^{\frac{s-s'-\lambda}{2}}\alpha^{a_1}\ldots
\alpha^{a_{s'}}}{s'!\sqrt{(\frac{s-s'+\lambda}{2})! (\frac{s-s'-\lambda}{2})!}}
\, \phibf_{\cur, \lambda}^{a_1\ldots a_{s'}} |0\rangle$
\\[2mm]
current in $R^{d-1,1}$& {\small $ s'=0,1,\ldots, s $} &
\\ [1mm]
& {\small $\lambda \in [s-s']_2$ }& \hspace{2cm} $(N_\alphabf + N_z+ N_\zeta-s)|\phibf\rangle=0$
\\ [1mm]\hline
&&
\\[-3mm]
anomalous spin-$s$ & $\phibf_{\sh, \lambda}^{a_1\ldots a_{s'}}$  & $
\sum\limits_{s'=0}^s\,\,\sum\limits_{\lambda \in [s-s']_2}
\frac{\zeta_{\phantom{z}}^{\frac{s-s'-\lambda}{2}}
\alpha_z^{\frac{s-s'+\lambda}{2}}\alpha^{a_1}\ldots
\alpha^{a_{s'}}}{s'!\sqrt{(\frac{s-s'+\lambda}{2})! (\frac{s-s'-\lambda}{2})!}}
\, \phibf_{\sh, \lambda}^{a_1\ldots a_{s'}} |0\rangle$
\\[2mm]
shadow in $R^{d-1,1}$& {\small $ s'=0,1,\ldots, s $} &
\\ [1mm]
& {\small $\lambda \in [s-s']_2$ }& \hspace{2cm} $(N_\alphabf + N_z+ N_\zeta-s)|\phibf\rangle=0$
\\ [1mm]\hline
\end{tabular}
\end{center}

\newpage
\noindent{\bf v}) Ordinary-derivative Lagrangian description of conformal spin-$s$ field
in $R^{d-1,1}$ with the field content given in Table I was discussed in
Refs.\cite{Metsaev:2007fq,Metsaev:2007rw}.%
\footnote{ In framework of AdS/CFT correspondence, recent discussion of conformal fields may be found in Refs.\cite{Giombi:2013yva,Tseytlin:2013jya}.}

\noindent {\bf Action and Lagrangian}. Gauge invariant action for fields in
flat and AdS spaces we found is given by
\beq
&&  S  =  \int d^dx\,  \LL \,,  \hspace{3.9cm}
\hbox{ for  fields  in }  \  R^{d-1,1},
\nonumber\\[-9pt]
\label{man02-26112011-10} &&
\\[-9pt]
&&  S   =  \int  d^dx\,  dz\,   \LL \,,
\hspace{3.4cm}  \hbox{  for  fields  in}  \  AdS_{d+1},
\nonumber
\eeq
where the Lagrangian is given by
\beq
\label{man02-26112011-11xx}
&& \LL  =    \half  \langle  \phibf|\mubf (\Box - \MM^2) |\phibf\rangle  +  \half  \langle  \bar\Lbf\phibf| |\bar\Lbf \phibf \rangle\,,
\\
\label{man02-26112011-12} && \hspace{1.3cm}   \bar\Lbf  \equiv
\bar\alphabf \partialbf  -  \half \alphabf \partialbf  \bar\alphabf^2 -
\eb_1 \Pibf^\smponetwo  +  \half e_1  \bar\alphabf^2\,,
\eeq
$\Box \equiv \partial^a\partial^a$, $|\bar \Lbf \phibf\rangle \equiv \bar\Lbf |\phibf\rangle$ and expressions $\alphabf\partialbf$, $\alphabf^2$, $ \mubf$, $\Pibf^\smponetwo$ are defined
in Appendix. The bra-vectors $\langle\phibf|$, $\langle \bar\Lbf \phibf|$ are defined as follows $\langle\phibf| \equiv (|\phibf\rangle)^\dagger$, $\langle \bar\Lbf \phibf| \equiv (|\bar\Lbf \phibf\rangle)^\dagger$. Operators $\MM^2$, $e_1$, $\eb_1$ appearing in \rf{man02-26112011-11xx},\rf{man02-26112011-12} are referred to as ladder operators in this paper. Explicit expressions for the ladder operators are given in Table II.
From \rf{man02-26112011-11xx},\rf{man02-26112011-12} and Table II, we see that Lagrangians for various fields are distinguished only by the ladder operators.

The following remarks are in order.

\noindent {\bf i}) We refer to the quantity $\bar\Lbf|\phibf\rangle$
as modified de Donder divergence. From Table II, we see that, only for massless field in flat space, $e_1=0$, $\eb_1=0$. This implies that, only for
massless field in flat space, the $\bar\Lbf|\phibf\rangle$
coincides with the standard de Donder divergence. From
\rf{man02-26112011-11xx},\rf{man02-26112011-12}, it is clear that many complicated terms contributing to the Lagrangian are collected into
$\langle\bar\Lbf\phibf|\bar\Lbf\phibf\rangle$-term. Thus, we see that it the use of
the modified de Donder divergence that allows us to
simplify significantly a structure of the Lagrangian.%
\footnote{ Applications of the standard de Donder-Feynman gauge condition to the
problems of higher-spin gauge fields may be found in
Refs.\cite{Guttenberg:2008qe}-\cite{Manvelyan:2008ks}. Recent discussion of
modified de Donder gauge may be found in Ref.\cite{Chang:2011mz}.}

\noindent {\bf ii}) Representation for Lagrangian in \rf{man02-26112011-11xx} is
valid for all theories whose symmetry algebras involve the Poincar\'e algebra. The corresponding Lagrangians are distinguished by the ladder operators $\MM^2$,
$e_1$, $\eb_1$. Namely, from \rf{man02-26112011-11xx}, we see that the dependence of the kinetic operator on the oscillators $\alpha^a$, $\bar\alpha^a$ and the
flat derivative $\partial^a$ takes the same form for massless and
massive fields in flat and AdS spaces and conformal fields in flat
space. In other words, the kinetic operators for just mentioned fields are distinguished only by the ladder operators $\MM^2$, $e_1$, and $\eb_1$. Thus we see that it is use of the ladder operators that allow us to treat AdS fields and conformal fields on an equal footing.

\noindent {\bf iii}) For massive field in flat space, representation of the gauge
invariant Lagrangian  in terms of modified de Donder divergence \rf{man02-26112011-11xx} was obtained in Ref.\cite{Metsaev:2008fs}, while, for massless and massive fields in AdS space,
such representation for Lagrangian was obtained in Refs.\cite{Metsaev:2008ks,Metsaev:2009hp}. For conformal field in flat space, Lagrangian \rf{man02-26112011-11xx} was obtained in Refs.\cite{Metsaev:2007fq,Metsaev:2007rw}.

{\bf Gauge symmetries}. We now discuss gauge symmetries of Lagrangian given
in \rf{man02-26112011-11xx}. Gauge
transformation parameters involved in gauge transformations of gauge fields are presented in Table III. We note that all gauge transformation parameters are traceless totally symmetric tensors of the Lorentz algebra $so(d-1,1)$.

The following remarks are in order.

\noindent{\bf i}) Gauge symmetries of massless spin-$s$ field
in $R^{d-1,1}$ with the gauge transformation parameter given in Table III were discussed in Ref.\cite{Fronsdal:1978rb}.

\noindent{\bf ii}) Gauge symmetries of massive spin-$s$ field in $R^{d-1,1}$ with the set of gauge transformations parameters given in Table III were discussed in Ref.\cite{Zinoviev:2001dt}.

\noindent {\bf iii}) Gauge symmetries of massless spin-$s$ field in
$AdS_{d+1}$ with set of gauge transformation parameters given in Table III were discussed
in Ref.\cite{Metsaev:2008ks}. This is to say that in Ref.\cite{Fronsdal:1978vb}, gauge symmetries of massless field in $AdS_{d+1}$, $d=3$, are described by gauge transformation parameter
that is totally symmetric traceless tensor field of the Lorentz
algebra $so(d,1)$. The $so(d-1,1)$ tensorial components of this gauge
transformation parameter are not traceless. Gauge transformation parameter in
Ref.\cite{Fronsdal:1978vb} is related to our gauge transformation parameters given in Table III by invertible transformation described in Ref.\cite{Metsaev:2008ks}.

\noindent {\bf iv})  Gauge symmetries of massive spin-$s$ field in $AdS_{d+1}$ with the set of gauge transformation parameters given in Table III were discussed in Ref.\cite{Metsaev:2009hp}. This is to say that in Ref.\cite{Zinoviev:2001dt}, gauge symmetries of massive field in $AdS_{d+1}$ are described by gauge transformation parameters that are totally symmetric traceless
tensor fields of the Lorentz algebra $so(d,1)$. The $so(d-1,1)$
tensorial components of the gauge transformation parameters in Ref.\cite{Zinoviev:2001dt} are
not traceless. Gauge transformation parameters in
Ref.\cite{Zinoviev:2001dt} are related to our gauge transformation
parameters given in Table III by invertible transformation
described in Ref.\cite{Metsaev:2009hp}.

\noindent {\bf v})  Gauge symmetries of conformal spin-$s$
field  in $R^{d-1,1}$ with the set of gauge transformation parameters given in Table III were introduced
in Refs.\cite{Metsaev:2007fq,Metsaev:2007rw}.

Using representation of the field contents and gauge transformation
parameters in terms of the ket-vectors $|\phibf\rangle$ and
$|\xibf\rangle$ we now note that the gauge transformations can entirely
be presented in terms of these ket-vectors. This is to say that the representation for
gauge transformations found in
Refs.\cite{Metsaev:2008ks}-\cite{Metsaev:2007rw} is given by
\be \label{man02-26112011-23}
\delta |\phibf\rangle = G |\xibf\rangle \,,
\qquad G \equiv \alphabf\partialbf - e_1 - \alphabf^2 \frac{1}{2N_\alphabf
+d-2}\eb_1 \,,
\ee
where the ladder operators $e_1$ and $\eb_1$ are given in Table II. For massless and massive fields in $R^{d-1,1}$, gauge
transformations in \rf{man02-26112011-23} coincide with the
ones in Refs.\cite{Fronsdal:1978rb,Zinoviev:2001dt}. For massless and massive fields in $AdS_{d+1}$, the gauge transformations in
Refs.\cite{Fronsdal:1978vb,Zinoviev:2001dt} can be cast into the form given in
\rf{man02-26112011-23} (see Refs.\cite{Metsaev:2008ks,Metsaev:2009hp}).

Summarizing the discussion of gauge invariant formulation of fields, we note that our gauge invariant formulation allows us to demonstrate explicitly how the ladder operators $e_1$, $\eb_1$ appearing in the gauge invariant Lagrangian \rf{man02-26112011-12} manifest themselves in the gauge transformations \rf{man02-26112011-23}.

\newpage{\small
\noindent {\sf Table II. For fields, the ladder operators $\MM^2$, $e_1$, $\eb_1$ enter Lagrangian  \rf{man02-26112011-11xx} and gauge transformation \rf{man02-26112011-23}. For currents and shadows, the ladder operators $e_1$, $\eb_1$ enter differential constraint \rf{man02-12122013-03} and gauge transformation \rf{man02-15122013-01}. In Table, $\m$ is a mass parameter of massive fields.}}
\begin{center}
\begin{tabular}{|l|c|c|c|}
\hline &&&
\\[-3mm]
Fields   & $\MM^2$  & $e_1$  &  $\eb_1$
\\ [1mm]\hline
&&&
\\[-1mm]
massless spin-$s$ & 0  & 0  &   0
\\[0mm]
field in $R^{d-1,1}$ &&&
\\[2mm]\hline
&&&
\\[-3mm]
massive spin-$s$ & &   &
\\[0mm]
field in $R^{d-1,1}$ & $\m^2$ & $ \m \zeta e_\zeta  $ & $ - \m e_\zeta \bar\zeta  $
\\[2mm]\hline
&  & &
\\[0mm]
massless spin-$s$ &  $ - \partial_z^2 + \frac{1}{z^2} (\nu^2-\frac{1}{4})$ & $ - \alpha^z e_z \TT_{\nu-\half}$ & $ -  \TT_{-\nu + \half} e_z \bar\alpha^z$
\\[3mm] \cline{2-4}\\[-5mm]
field in $AdS_{d+1}$ & \mc{3}{|c|}{} \\
& \mc{3}{|c|}{ $\nu \equiv s+\frac{d-4}{2} - N_z$, \qquad $\TT_\nu\equiv \partial_z +\frac{\nu}{z}$ }
\\[2mm]\hline
&  &   &
\\[0mm]
massive spin-$s$ & $- \partial_z^2 + \frac{1}{z^2} (\nu^2-\frac{1}{4})$ & $ - \zeta r_\zeta \TT_{ -\nu - \half} - \alpha^z r_z \TT_{\nu-\half}$ & $
- \TT_{\nu + \half}  r_\zeta \bar\zeta  - \TT_{-\nu + \half} r_z
\bar\alpha^z $
\\[3mm] \cline{2-4}\\[-5mm]
field in $AdS_{d+1}$ & \mc{3}{|c|}{} \\
& \mc{3}{|c|}{ $\nu \equiv \kappa + N_\zeta - N_z$, \qquad $\TT_\nu\equiv \partial_z +\frac{\nu}{z}$  }
\\[2mm]\hline
conformal spin-$s$ & & &
\\[2mm]
field in $R^{d-1,1}$ &  $ \upsilon^\oplussm \bar\upsilon^\oplussm$ & $ \zeta e_\zeta \bar\upsilon^\oplussm $ &  $ - \upsilon^\oplussm e_\zeta \bar\zeta $
\\[3mm] \hline
canonical spin-$s$ &   &   &
\\[0mm]
current in $R^{d-1,1}$ & - & $ \alpha^z e_z $  & $ - e_z \bar\alpha^z \Box $
\\[3mm] \hline
canonical spin-$s$ &   &   &
\\[0mm]
shadow in $R^{d-1,1}$ & - & $ \alpha^z e_z \Box $  & $ - e_z \bar\alpha^z  $
\\[2mm]\hline
anomalous spin-$s$ &   &   &
\\[0mm]
current in $R^{d-1,1}$ & - & $ \zeta r_\zeta \Box + \alpha^zr_z $  & $ -r_\zeta\bar\zeta  - r_z \bar\alpha^z \Box $
\\[2mm]\hline
anomalous spin-$s$ &   &   &
\\[0mm]
shadow in $R^{d-1,1}$ & - & $ \zeta r_\zeta  + \alpha^zr_z \Box $  & $ -r_\zeta\bar\zeta \Box - r_z \bar\alpha^z  $
\\[2mm]\hline
\mc{4}{|c|}{}
\\
\mc{4}{|c|}{$ e_\zeta = \Bigl(\frac{2s+d-4-N_\zeta}{2s+d-4-2N_\zeta}\Bigr)^{1/2}$, \qquad $e_z = \Bigl(\frac{2s+d-4-N_z}{2s+d-4-2N_z}\Bigr)^{1/2}$}
\\
\mc{4}{|c|}{}
\\
\mc{4}{|c|}{ $ r_\zeta = \left(\frac{(s+\frac{d-4}{2}
-N_\zeta)(\kappa - s-\frac{d-4}{2} + N_\zeta)(\kappa + 1 +
N_\zeta)}{2(s+\frac{d-4}{2}-N_\zeta - N_z)(\kappa +N_\zeta -N_z) (\kappa+
N_\zeta - N_z +1)}\right)^{1/2} $ }
\\
\mc{4}{|c|}{}
\\
\mc{4}{|c|}{ $ r_z = \left(\frac{(s+\frac{d-4}{2} -N_z)(\kappa
+ s + \frac{d-4}{2} - N_z)(\kappa - 1 -
N_z)}{2(s+\frac{d-4}{2}-N_\zeta-N_z)(\kappa + N_\zeta - N_z) (\kappa +N_\zeta
- N_z -1)}\right)^{1/2} $ }
\\
\mc{4}{|c|}{}
\\
\mc{4}{|c|}{ $N_\zeta = \zeta\bar\zeta$, \qquad $N_z = \alpha^z\bar\alpha^z$ }
\\
\mc{4}{|c|}{}
\\
\mc{4}{|c|}{ $ \kappa \equiv \sqrt{\m^2 + \Bigl( s+
\frac{d-4}{2}\Bigr)^2} $ }
\\ [3mm]\hline
\end{tabular}
\end{center}

\newpage
{\small
\noindent {\sf Table III. Gauge transformation parameters and the corresponding ket-vectors entering gauge transformation of fields, currents, and shadows. Gauge transformation parameters with $s'\geq 2$ are totally symmetric traceless tensor fields of $so(d-1,1)$ algebra, $\xibf^{aaa_3\ldots a_{s'}}=0$, $s'\geq 2$.}}
\begin{center}
\begin{tabular}{|l|c|l|}
\hline &&
\\[-3mm]
Fields   & Gauge transformation  & Ket-vector $|\xibf\rangle$
\\ [1mm]
& parameters  &
\\ [1mm]\hline
&&
\\[-1mm]
massless spin-$s$ & $ \xibf^{a_1\ldots a_{s-1}} $  & $\frac{1}{(s-1)!} \alpha^{a_1} \ldots \alpha^{a_{s-1}}
\xibf^{a_1\ldots a_{s-1}} |0\rangle $
\\[0mm]
field in $R^{d-1,1}$ &&
\\ [1mm]\hline
&&
\\[-1mm]
massive spin-$s$ & $\xibf^{a_1\ldots a_{s'}}$  & $\sum\limits_{s'=0}^{s-1} \frac{\zeta^{s-1-s'} \alpha^{a_1} \ldots
\alpha^{a_{s'}}}{s'!\sqrt{(s -1 - s')!}} \, \xibf^{a_1\ldots a_{s'}}
|0\rangle$
\\[-3mm]
field in $R^{d-1,1}$ & \quad {\small $  s'=0,1,\ldots,s-1 $} &
\\ [1mm]\hline
&&
\\[-1mm]
massless spin-$s$ & $\xibf^{a_1\ldots a_{s'}} $  & $ \sum\limits_{s'=0}^{s-1} \frac{\alpha_z^{s-1-s'} \alpha^{a_1} \ldots
\alpha^{a_{s'}}}{s'!\sqrt{(s-1 - s')!}} \, \xibf^{a_1\ldots a_{s'}}
|0\rangle $
\\[-1mm]
field in $AdS_{d+1} $ & {\small $  s'=0,1,\ldots,s-1 $} &
\\ [1mm]\hline
&&
\\[-1mm]
massive spin-$s$ & $ \xibf_\lambda^{a_1\ldots a_{s'}} $  & $\sum\limits_{s'=0}^{s-1} \, \, \sum\limits_{\lambda\in [s-1-s']_2}
\frac{\zeta_{\phantom{z}}^{\frac{s-1-s'+ \lambda}{2}}
\alpha_z^{\frac{s-1-s'- \lambda}{2}}\alpha^{a_1}\ldots
\alpha^{a_{s'}}}{s'!\sqrt{(\frac{s-1-s'+ \lambda}{2})!
(\frac{s-1-s'- \lambda}{2})!}} \, \xibf_\lambda^{a_1\ldots a_{s'}} |0\rangle $
\\[-2mm]
field in $AdS_{d+1} $ & \quad {\small $  s'=0,1,\ldots,s-1 $}  &
\\ [1mm]
& {\small $\lambda \in [s-1-s']_2$} &
\\ [1mm]\hline
&&
\\[-1mm]
conformal spin-$s$ & $ \xibf_{k'-1}^{a_1\ldots a_{s'}} $  & $\sum\limits_{s'=0}^{s-1} \sum\limits_{k'\in [k_{s'}+1]_2}
\frac{\zeta^{s-1-s'} (\upsilon^\ominussm)^{^{\frac{k_{s'}+1+k'}{2}}}
(\upsilon^\oplussm)^{^{\frac{k_{s'}+1-k'}{2}}}}{ s'!
\sqrt{(s-1-s')!} (\frac{k_{s'}+1+k'}{2})!} $
\\[-2mm]
field in $R^{d-1,1}$ & {\small $  s'=0,1,\ldots,s-1 $} &
\\ [1mm]
& {\small $ k' \in [k_{s'}+1]_2$ }  & $ \hspace{2.3cm} \times \ \alpha^{a_1} \ldots
\alpha^{a_{s'}} \, \xibf_{k'-1}^{a_1\ldots a_{s'}}
|0\rangle $
\\ [1mm]
& {\small $k_{s'} \equiv s' + \frac{d-6}{2}$ } &
\\ [1mm]\hline
&&
\\[-1mm]
canonical spin-$s$ & $\xibf_\cur^{a_1\ldots a_{s'}}$  & $\sum\limits_{s'=0}^{s-1} \frac{\alpha_z^{s-1-s'} \alpha^{a_1} \ldots
\alpha^{a_{s'}}}{s'!\sqrt{(s -1 - s')!}} \, \xibf_\cur^{a_1\ldots a_{s'}}
|0\rangle$
\\[-2mm]
current in $R^{d-1,1}$ & {\small $ s'=0,1,\ldots, s-1 $ }&
\\[1mm]\hline
&&
\\[-1mm]
canonical spin-$s$ & $\xibf_\sh^{a_1\ldots a_{s'}} $  & $ \sum\limits_{s'=0}^{s-1} \frac{\alpha_z^{s-1-s'} \alpha^{a_1} \ldots
\alpha^{a_{s'}}}{s'!\sqrt{(s-1 - s')!}} \, \xibf_\sh^{a_1\ldots a_{s'}}
|0\rangle $
\\[-2mm]
shadow in $R^{d-1,1}$ & {\small $ s'=0,1,\ldots, s-1 $ } &
\\ [1mm]\hline
&&
\\[-2mm]
anomalous spin-$s$ & $ \xibf_{\cur,\lambda}^{a_1\ldots a_{s'}} $  &
\\[-2mm]
current in $R^{d-1,1}$ & {\small $ s'=0,1,\ldots, s-1 $ } & $\sum\limits_{s'=0}^{s-1} \, \, \sum\limits_{\lambda\in [s-1-s']_2}
\frac{\zeta_{\phantom{z}}^{\frac{s-1-s'+\lambda}{2}}
\alpha_z^{\frac{s-1-s'-\lambda}{2}}\alpha^{a_1}\ldots
\alpha^{a_{s'}}}{s'!\sqrt{(\frac{s-1-s'+\lambda}{2})!
(\frac{s-1-s'-\lambda}{2})!}} \, \xibf_{\cur,\lambda}^{a_1\ldots a_{s'}} |0\rangle $
\\[-2mm]
& {\small $\lambda\in [s-1-s']_2$ }&
\\[2mm]\hline
&&
\\[-2mm]
anomalous spin-$s$ & $ \xibf_{\sh,\lambda}^{a_1\ldots a_{s'}} $  &
\\[-2mm]
shadow in $R^{d-1,1}$ & {\small $ s'=0,1,\ldots, s-1 $ } & $\sum\limits_{s'=0}^{s-1} \, \, \sum\limits_{\lambda\in [s-1-s']_2}
\frac{\zeta_{\phantom{z}}^{\frac{s-1-s'-\lambda}{2}}
\alpha_z^{\frac{s-1-s'+\lambda}{2}}\alpha^{a_1}\ldots
\alpha^{a_{s'}}}{s'!\sqrt{(\frac{s-1-s'+\lambda}{2})!
(\frac{s-1-s'-\lambda}{2})!}} \, \xibf_{\sh,\lambda}^{a_1\ldots a_{s'}} |0\rangle $
\\[-2mm]
& {\small $\lambda\in [s-1-s']_2$ }&
\\[1mm]\hline
\end{tabular}
\end{center}

\newsection{ \large Gauge invariant approach to currents and shadows
} \label{section-03}

As we obtain our light-cone gauge approach to currents and shadows by using gauge invariant approach developed in Refs.\cite{Metsaev:2008fs}-\cite{Metsaev:2011uy}, we now review the gauge invariant approach in Refs.\cite{Metsaev:2008fs}-\cite{Metsaev:2011uy}.

Let us start with a brief recall of some basic notions of CFT. Fields of CFT can be separated into two groups: currents and shadows.
In this paper, field having Lorentz algebra spin $s$, $s\geq 1$, and conformal dimension
$\Delta = s+d-2$ is referred to as canonical current, while field
having Lorentz algebra spin $s$, $s\geq 1$, and conformal dimension $\Delta > s+d-2$ is
referred to as anomalous current. Accordingly, field having Lorentz algebra spin
$s$, $s\geq 1$, and conformal dimension
$\Delta = 2 - s$ is referred to as canonical shadow,
while field having Lorentz algebra spin $s$, $s\geq 1$, and conformal
dimension $\Delta < 2 - s$ is referred to as anomalous shadow.

In Refs.\cite{Metsaev:2008fs,Metsaev:2009ym}, we developed the gauge invariant formulation of the canonical currents and shadows, while, in Refs.\cite{Metsaev:2010zu,Metsaev:2011uy},
we generalized our gauge invariant approach to the case of anomalous currents and shadows.%
\footnote{ Before our discussion in Refs.\cite{Metsaev:2010zu,Metsaev:2011uy}, gauge invariant approach to anomalous currents was studied in Ref.\cite{Gover:2008sw}. In Ref.\cite{Gover:2008sw}, the gauge invariant approach was developed by using the tractor approach. Discussion of various aspects of the tractor approach may be found in Refs.\cite{Gover:2008pt}.}
Our gauge invariant approach to currents and shadows can be summarized as follows.

\noindent {\bf i}) Starting with a field content of currents (shadows) appearing in the standard CFT, we introduce auxiliary fields and Stueckelberg fields, i.e., we extend space of fields entering the standard CFT.

\noindent {\bf ii}) On the extended space of fields, we introduce
differential constraints, gauge transformations, and $so(d,2)$ algebra
transformations. The differential constraints are required to be invariant
under the gauge transformations and the $so(d,2)$ algebra transformations.%
\footnote{ For the discussion of
differential constraints for conformal currents in the framework of standard CFT see, e.g., Refs.\cite{Shaynkman:2004vu}.}

\noindent {\bf iii})  The gauge symmetries and the differential constraints allow us to match our approach and the standard CFT, i.e., by gauging away the Stueckelberg fields and by solving differential constraints to exclude the auxiliary fields, we get formulation of currents and shadows in the standard CFT.

We now start our brief review of  our gauge invariant approach to currents and
shadows with the discussion of field contents.

\noindent {\bf Field content of spin-$s$ canonical current and spin-$s$ canonical shadow}. To discuss the gauge invariant formulation of arbitrary spin-$s$ canonical current and spin-$s$
canonical shadow we use the respective totally symmetric $so(d-1,1)$ Lorentz
algebra tensor fields which are assumed to be double-traceless,
\beq
\label{man02-12122013-01}
&& \phibf_\cur^{a_1\ldots a_{s'}}\,, \hspace{2.2cm} \phibf_\sh^{a_1\ldots
a_{s'}}\,, \hspace{2.2cm}  s'=0,1,\ldots,s\,,
\\
\label{man02-15072009-02} && \phibf_\cur^{aabba_5\ldots a_{s'}}=0\,, \qquad
\phibf_\sh^{aabba_5\ldots a_{s'}}=0\,, \qquad \hbox{ for} \ s' \geq 4\,.
\eeq
Conformal dimensions of the fields in \rf{man02-12122013-01} are given by
\be
\Delta(\phibf_\cur^{a_1\ldots a_{s'}}) = s'+d-2\,, \qquad \Delta(\phibf_\sh^{a_1\ldots
a_{s'}}) = 2-s'.
\ee

\noindent {\bf Field content of spin-$s$ anomalous current and spin-$s$ anomalous shadow}. To discuss the gauge invariant formulation of arbitrary spin-$s$ anomalous current and spin-$s$
anomalous shadow we use the respective totally symmetric $so(d-1,1)$ Lorentz
algebra tensor fields which are assumed to be double-traceless,
\beq
\label{man02-12122013-02}
&& \phibf_{\cur,\lambda}^{a_1\ldots a_{s'}}\,, \hspace{2.2cm} \phibf_{\sh,\lambda}^{a_1\ldots
a_{s'}}\,, \qquad  s'=0,1,\ldots,s\,,\qquad \lambda \in [s-s']_2,\qquad
\\
\label{man02-15072009-02x} && \phibf_{\cur,\lambda}^{aabba_5\ldots a_{s'}}=0\,, \qquad
\phibf_{\sh,\lambda}^{aabba_5\ldots a_{s'}}=0\,, \qquad \hbox{ for} \ s' \geq 4\,.
\eeq
Conformal dimensions of the fields in \rf{man02-12122013-02} are given by
\be \label{man02-15072009-02xx}
\Delta(\phibf_{\cur,\lambda}^{a_1\ldots a_{s'}})= \frac{d}{2} + \kappa +
\lambda\,,\qquad
\Delta(\phibf_{\sh,\lambda}^{a_1\ldots a_{s'}})= \frac{d}{2} - \kappa +
\lambda\,.
\ee
Summary of the field contents we use and appropriate ket-vectors are given in Table I.
As shown in Refs.\cite{Metsaev:2010zu,Metsaev:2011uy}, in the framework of AdS/CFT correspondence, the parameter $\kappa$ in \rf{man02-15072009-02xx} is related to
the mass parameter $\m$ of spin-$s$ massive field in $AdS_{d+1}$ as in Table II.

{\bf Differential constraint for current and shadow}. For the ket-vectors of current and shadow given in Table I, we introduce the following differential constraint:
\beq
\label{man02-12122013-03} \bar\Lbf |\phibf\rangle & = & 0 \,,
\\
\label{man02-12122013-03x} && \bar\Lbf  = \bar\alphabf\partialbf - \half \alphabf\partialbf  \bar\alphabf^2
- \eb_1 \Pibf^\smponetwo + \half e_1 \bar\alphabf^2 \,,
\eeq
where the operators $e_1$ and $\eb_1$ are given in Table II, while the operator $\Pibf^\smponetwo$ is given in \rf{man02-17122013-02a}.
We note that constraint \rf{man02-12122013-03} is invariant under gauge transformation and
$so(d,2)$ algebra transformations which we discuss below.

\noindent {\bf Two-point gauge invariant vertices}. For currents and shadows, one can construct two gauge invariant 2-point vertices. The first 2-point vertex, denoted by $\Gamma^{\rm cur-\sh}$, is a local functional of current and shadow, while the second 2-point vertex, denoted by $\Gamma^{\rm sh-sh}$, is a non-local functional of shadows. Using notation $|\phibf_\cur\rangle$ and $|\phibf_\sh\rangle$ for the respective ket-vectors of currents and shadows given in Table I, we note the following expressions for the vertices:
\beq
\label{man02-15122013-07} && \Gamma^{\rm cur-\sh} = \int d^d x  \LL^{\rm cur-sh}\,, \qquad \LL^{\rm cur-sh} = \langle \phibf_\cur|\mubf |
\phibf_\sh\rangle \,,
\\
\label{man02-15122013-08} && \Gamma^{\rm sh-sh}= \int d^dx_1 d^dx_2 \LL_{12}^{\rm sh-sh} \,,
\\
&& \hspace{1.3cm} \LL_{12}^{\rm sh-sh} \equiv \half \langle\phibf_\sh(x_1)|
\frac{\mubf f_\nu}{ |x_{12}|^{2\nu + d }} |\phibf_\sh (x_2)\rangle \,,
\\
\label{man02-15122013-09} && \hspace{1.3cm} f_\nu \equiv \frac{\Gamma(\nu + \frac{d}{2})\Gamma(\nu + 1)}{4^{\bar\kappa - \nu}
\Gamma(\bar\kappa + \frac{d}{2})\Gamma(\bar\kappa + 1)} \,,
\\
&& \hspace{1.3cm} |x_{12}|^2 \equiv x_{12}^a x_{12}^a\,, \qquad x_{12}^a = x_1^a - x_2^a\,,
\eeq
\beq
\label{man02-15122013-10} &&\hspace{-0.5cm}  \nu \equiv s+ \frac{d-4}{2} - N_z, \qquad \bar\kappa \equiv  s + \frac{d-4}{2}, \hspace{1cm} \hbox{ for canonical shadow},\qquad
\\
\label{man02-15122013-11} && \hspace{-0.5cm} \nu \equiv \kappa + N_\zeta -N_z, \hspace{1.4cm} \bar\kappa \equiv \kappa, \hspace{2.4cm} \hbox{ for anomalous shadow},
\eeq
where operators $N_\zeta$, $N_z$, $\mubf$ are defined in Appendix.

\noindent {\bf Gauge transformations parameter for canonical current and shadow}. To discuss gauge symmetries of spin-$s$ canonical current and spin-$s$ canonical shadow we use the respective gauge transformation parameters
\be \label{man02-12122013-04}
\xibf_\cur^{a_1\ldots a_{s'}}\,, \qquad \xibf_\sh^{a_1\ldots
a_{s'}}\,, \qquad  s'=0,1,\ldots,s-1\,,
\ee
which are totally symmetric fields of the $so(d-1,1)$ Lorentz algebra. The parameters are assumed to be traceless,
\be \label{man02-15072009-05}
\xibf_\cur^{aaa_3\ldots a_{s'}}=0\,, \qquad
\xibf_\sh^{aaa_3\ldots a_{s'}}=0\,, \qquad \hbox{ for} \ s' \geq 2\,.
\ee
Conformal dimensions of the gauge transformation parameters in \rf{man02-12122013-04} are given by
\be
\Delta(\xibf_\cur^{a_1\ldots a_{s'}}) = s'+d-3\,, \qquad \Delta(\xibf_\sh^{a_1\ldots
a_{s'}}) = 1-s'.
\ee

\noindent {\bf Gauge transformations parameter for anomalous current and shadow}. To discuss gauge symmetries of spin-$s$ anomalous current and spin-$s$ anomalous shadow we use the respective gauge transformation parameters
\be \label{man02-12122013-05}
\xibf_{\cur,\lambda}^{a_1\ldots a_{s'}}\,, \qquad \xibf_{\sh,\lambda}^{a_1\ldots
a_{s'}}\,, \qquad  s'=0,1,\ldots,s-1\,,\qquad \lambda \in [s-1-s']_2\,,
\ee
which are totally symmetric fields of the $so(d-1,1)$ Lorentz algebra. The parameters are assumed to be traceless,
\be\label{man02-12122013-06}
\xibf_{\cur,\lambda}^{aaa_3\ldots a_{s'}}=0\,, \qquad
\xibf_{\sh,\lambda}^{aaa_3\ldots a_{s'}}=0\,, \qquad \hbox{ for} \ s' \geq 2\,.
\ee
Conformal dimensions of the gauge transformation parameters in \rf{man02-12122013-05} are given by
\be
\Delta(\xibf_{\cur,\lambda}^{a_1\ldots a_{s'}})= \frac{d}{2} + \kappa +
\lambda - 1\,,\qquad
\Delta(\xibf_{\sh,\lambda}^{a_1\ldots a_{s'}})= \frac{d}{2} - \kappa +
\lambda -1\,.
\ee
The gauge transformation parameters and the corresponding ket-vectors are given in Table III.

Using representation of the currents, shadows, and gauge transformation
parameters in terms of the ket-vectors $|\phibf\rangle$,
$|\xibf\rangle$ given in Tables I and III, we now note that the gauge transformations can entirely be presented in terms of these ket-vectors. This is to say that the gauge transformations found in Refs. \cite{Metsaev:2008fs}-\cite{Metsaev:2011uy} take the form
\be  \label{man02-15122013-01}
\delta |\phibf\rangle = G |\xibf\rangle \,,
\qquad G \equiv \alphabf\partialbf - e_1 - \alphabf^2 \frac{1}{2N_\alphabf
+d-2}\eb_1 \,,
\ee
where the ladder operators $e_1$ and $\eb_1$ are given in Table II. The following remarks are in order.

\noindent {\bf i}) Gauge transformation \rf{man02-15122013-01} takes the same form as the one in \rf{man02-26112011-23}, i.e., gauge transformation for fields in \rf{man02-26112011-23} is distinguished from the one for currents and shadows \rf{man02-15122013-01} only by the ladder operators $e_1$ and $\eb_1$.

\noindent {\bf ii}) Differential constraint \rf{man02-12122013-03} is invariant under gauge transformation \rf{man02-15122013-01}.

\noindent {\bf iii}) Relation of our approach to the standard CFT is achieved by using Stueckelberg gauge frame. This frame implies gauging away Stueckelberg fields and solving auxiliary  field via the differential constraint.
For the case of the spin-$s$ canonical current, the use of the Stueckelberg gauge frame leads to divergence-free and tracelessness constraints for $\phibf_\cur^{a_1\ldots a_s}$,
\be
\partial^a \phibf_\cur^{aa_2\ldots a_s} = 0 \,, \qquad \phibf_\cur^{aaa_3\ldots a_s}=
0\,,
\ee
i.e., we see that, in the Stueckelberg gauge frame, our field $\phibf_\cur^{a_1\ldots a_s}$ can be identified with the conserved current in the standard CFT.
For the case of the spin-$s$ anomalous current, the use of the Stueckelberg gauge frame leads to tracelessness constraint for $\phibf_\cur^{a_1\ldots a_s}$.
For the spin-$s$ shadow, the use of the Stueckelberg gauge frame leads to
tracelessness constraint for the field $\phibf_\sh^{a_1\ldots a_{s}}$,
which can be identified with the shadow
in the standard approach to CFT. For more detailed study of relation of our approach to the standard CFT,  see Refs.\cite{Metsaev:2008fs}-\cite{Metsaev:2011uy}.

\newsection{ \large Relativistic symmetries of fields, currents, and shadows }\label{section-04}

Relativistic symmetries of massless and massive fields in $R^{d-1,1}$ are described by the Poincar\'e symmetries, while relativistic symmetries of fields in $AdS_{d+1}$, conformal fields, currents, and shadows in $R^{d-1,1}$ are described by the $so(d,2)$
symmetries. In our approach, in order to treat fields, currents, and shadows on equal footing, only $so(d-1,1)$ symmetries are realized manifestly.
Therefore it is reasonable to represent the $so(d,2)$ algebra so that to
respect the manifest $so(d-1,1)$ symmetries.%
\footnote{ For the case of AdS fields, the $so(d,2)$ symmetries can manifestly be realized  by using ambient space approach (see e.g. Refs.\cite{Metsaev:1995re}-\cite{Fotopoulos:2006ci}).}
This is to say that the $so(d,2)$ algebra consists of translation generators $P^a$, conformal boost generators $K^a$, dilatation
generator $D$, and generators $J^{ab}$ which span $so(d-1,1)$ algebra. Commutations relations of the $so(d,2)$ algebra generators take the form,
\beq
&& {}[D,P^a]=-P^a\,, \hspace{2.5cm}  [P^a,J^{bc}]=\eta^{ab}P^c -\eta^{ac}P^b
\,,
\\
&& [D,K^a]=K^a\,, \hspace{2.7cm} [K^a,J^{bc}]=\eta^{ab}K^c - \eta^{ac}K^b\,,
\\
&& [P^a,K^b]=\eta^{ab}D - J^{ab}\,, \hspace{1.2cm}  [J^{ab},J^{ce}]=\eta^{bc}J^{ae}+3\hbox{ terms} \,.
\eeq
For the case of fields in $AdS_{d+1}$ and conformal fields in $R^{d-1,1}$, requiring  the $so(d,2)$ symmetries implies that the actions of AdS fields and conformal fields \rf{man02-26112011-10} are invariant under the transformation $\delta |\phibf\rangle  =  G_\diff |\phibf\rangle$, while for the case of currents and shadows, requiring the $so(d,2)$ symmetries implies that differential constraints \rf{man02-12122013-03} are invariant under the transformation $\delta |\phibf\rangle  = G_\diff |\phibf\rangle$. For all these cases, the realization of $so(d,2)$ algebra generators $ G_\diff $ in terms of
differential operators acting on the respective ket-vectors $|\phibf\rangle$ takes the form
\beq
\label{conalggenlis01} && P^a = \partial^a \,,
\qquad
J^{ab} = x^a\partial^b -  x^b\partial^a + M^{ab}\,,
\\
\label{conalggenlis03} && D = x^a\partial^a  + \Delta\,,
\\
\label{conalggenlis04} && K^a = -\frac{1}{2}x^b x^b\partial^a + x^a D + M^{ab}x^b
+ R^a \,,
\\
\label{conalggenlis04x} && \hspace{1cm} M^{ab} = \alpha^a \bar\alpha^b - \alpha^b \bar\alpha^a\,.
\eeq
In \rf{conalggenlis01}-\rf{conalggenlis04x}, $M^{ab}$ is a spin operator of the
$so(d-1,1)$ algebra, while $\Delta$ in \rf{conalggenlis03} is a operator which, for the case of conformal field, currents, and shadows, is the operator of conformal dimension.
Operator $R^a$ in \rf{conalggenlis04} is independent of coordinates $x^a$. Explicit expressions for the operators $\Delta$ and $R^a$ are given in Table IV.

\newsection{ \large Light-cone gauge action of fields and 2-point vertices of currents and shadows } \label{section-05}

To develop light-cone gauge formulation of fields, currents, and shadows we use totally symmetric traceless tensor fields  of the $so(d-2)$ algebra. The field contents we use are presented in Table V. To simplify the presentation we use oscillators $\alpha^i$, $\alpha^z$, $\zeta$, $\upsilon^\oplussm$, $\upsilon^\ominussm$, $i=1,\ldots,d-2$,  and introduce the corresponding ket-vectors which are also presented in Table V. Note that, in terms  of the ket-vectors given in Table V, tracelessness constraint takes the form
\be \label{man02-18122013-01}
\bar\alpha^i \bar\alpha^i \phik = 0\,.
\ee
Also, note that, in light-cone gauge approach, ket-vectors $\phik$ of fields, currents, and shadows are not subject to differential constraint and should satisfy
only algebraic tracelessness constraint \rf{man02-18122013-01}.

\bigskip
\bigskip
{\small
\noindent {\sf Table IV. Operators $\Delta$ and $R^a$ entering
$so(d,2)$ transformations of AdS fields, conformal fields, currents, and shadows \rf{conalggenlis01}-\rf{conalggenlis04}. Operators $\widetilde\Abf^a$, $\bar\Abf_\perp^a$, $N_\alphabf$, $N_z$, $N_\zeta$, $N_{\upsilon^\oplussm}$, $N_{\upsilon^\ominussm}$ are defined in Appendix. Operators $e_z$, $e_\zeta$,  $r_z$, $r_\zeta$, and parameter $\kappa$ take the same form as in Table II.}
\begin{center}
\begin{tabular}{|l|c|l|}
\hline &&
\\[-3mm]
Fields   & Operator  $\Delta$ & Operator $R^a$
\\ [1mm]\hline
&&
\\[-1mm]
massless spin-$s$  & $ z\partial_z + \frac{d-1}{2} $  & $
r_{0,1} \bar\alpha^a + \rb_{0,1} \widetilde\Abf^a + r_{1,1}\partial^a \,,$
\\[3mm]
field in $AdS_{d+1}$&  & $r_{0,1} \equiv - z \alpha^z e_z\,, \quad \rb_{0,1} \equiv z  e_z \bar\alpha^z\,, \quad r_{1,1} \equiv - \half z^2$
\\ [2mm]\hline
&&
\\[-1mm]
massive spin-$s$  & $ z\partial_z + \frac{d-1}{2} $  & $
r_{0,1} \bar\alpha^a + \rb_{0,1} \widetilde\Abf^a + r_{1,1}\partial^a\,, $
\\[3mm]
field in $AdS_{d+1}$ & & $r_{0,1} \equiv - z (\zeta r_\zeta +\alpha^z r_z )\,, \quad \rb_{0,1} \equiv z ( r_\zeta \bar\zeta + r_z \bar\alpha^z)\,, $
\\ [2mm]
& & $ r_{1,1} \equiv  - \half z^2 $
\\ [1mm]\hline
&&
\\[-1mm]
conformal spin-$s$  & $ \frac{d-2}{2} - N_{\upsilon^\oplussm} + N_{\upsilon^\ominussm}  $  & $
r_{0,1} \bar\alpha^a + \rb_{0,1} \widetilde\Abf^a + r_{1,1}\partial^a\,, $
\\[3mm]
field in $R^{d-1,1}$ &  & $r_{0,1} \equiv  2 \zeta e_\zeta \bar\upsilon^\ominussm\,, \quad \rb_{0,1} \equiv - 2 \upsilon^\ominussm e_\zeta \bar\zeta\,, \quad r_{1,1} \equiv -2\upsilon^\ominussm \bar\upsilon^\ominussm $
\\ [2mm]\hline
&&
\\[-1mm]
canonical spin-$s$ & $ s+d-2-N_z $  & $
\rb \Bigl( \widetilde\Abf^a + \alphabf^2 \frac{2}{(2N_\alphabf + d - 2)(2N_\alphabf+d)} \bar\Abf_\perp^a\Bigr), $
\\[4mm]
current in $R^{d-1,1}$ &  & $\rb \equiv - \sqrt{(2s+d-4-N_z)(2s+d-4-2N_z)}\bar\alpha^z$
\\ [1mm]\hline
&&
\\[-1mm]
canonical spin-$s$ & $2-s+ N_z$  & $
r \Bigl( \bar\alpha^a - \alpha^a \frac{1}{2N_\alphabf+d } \bar\alphabf^2\Bigr), $
\\[5mm]
shadow in $R^{d-1,1}$  &   & $ r \equiv \alpha^z \sqrt{(2s+d-4-N_z)(2s+d-4-2N_z)}$
\\ [1mm]\hline
&&
\\[-1mm]
anomalous spin-$s$ & $ \frac{d}{2} + \kappa + N_\zeta - N_z $  & $
- 2 \zeta r_\zeta \Bigl( (\nu+1) \bar\alpha^a  -
\bar\Abf_\perp^a\Bigr) - 2\Bigl( \nu \widetilde\Abf^a + \alphabf^2 \frac{1}{2N_\alphabf+d-2} \bar\Abf_\perp^a
\Bigr)r_z\bar\alpha^z, $
\\[3mm]
current in $R^{d-1,1}$ & & $\nu \equiv \kappa + N_\zeta - N_z$
\\ [1mm]\hline
&&
\\[-1mm]
anomalous spin-$s$ & $  \frac{d}{2} - \kappa - N_\zeta + N_z  $  & $
 2 \alpha^z r_z \Bigl( (\nu-1) \bar\alpha^a  +
\bar\Abf_\perp^a\Bigr) +  2\Bigl( \nu \widetilde\Abf^a - \alphabf^2 \frac{1}{2N_\alphabf+d-2}
\bar\Abf_\perp^a \Bigr)r_\zeta\bar\zeta, $
\\[3mm]
shadow in $R^{d-1,1}$ &   & $\nu \equiv \kappa + N_\zeta - N_z$
\\ [2mm]\hline
\end{tabular}
\end{center}
}

\newpage
{\small
\noindent {\sf Table V. Field contents and the corresponding ket-vectors entering  the light-cone gauge formulation of fields, currents, and shadows. Fields with $s'\geq 2$ are totally symmetric traceless tensor fields of the $so(d-2)$ algebra, $\phi^{iii_3\ldots i_{s'}}=0$, $s'\geq 2$. i.e., $\bar\alpha^2 \phik = 0$.  For notion of $\lambda \in [s-s']_2$ and oscillator algebra see Appendix. In the Table, algebraic constraints express the homogeneity properties of ket-vectors $\phik$ with respect to the oscillators.}}
\begin{center}
\begin{tabular}{|l|c|l|}
\hline &&
\\[-3mm]
 & Field content   & Ket-vector $|\phi\rangle$ and algebraic constraints
\\ [1mm]\hline
&&
\\[-1mm]
massless spin-$s$ & $\phi^{i_1\ldots i_s}$  & $\frac{1}{s!} \alpha^{i_1} \ldots \alpha^{i_s}
\phi^{i_1\ldots i_s} |0\rangle$, \hspace{1.5cm} $(N_\alpha-s)\phik=0$
\\[1mm]
field in $R^{d-1,1}$ &&
\\ [1mm]\hline
&&
\\[-3mm]
massive spin-$s$ & $\phi^{i_1\ldots i_{s'}}$  & $
\sum\limits_{s'=0}^s \frac{\zeta^{s-s'} \alpha^{i_1} \ldots
\alpha^{i_{s'}}}{s'!\sqrt{(s - s')!}} \, \phi^{i_1\ldots i_{s'}} |0\rangle$,\qquad  $(N_\alpha+N_\zeta-s)\phik=0$
\\[2mm]
field in $R^{d-1,1}$ & {\small $ s'=0,1,\ldots, s $} &
\\ [1mm]\hline
&&
\\[-3mm]
massless spin-$s$ & $\phi^{i_1\ldots i_{s'}}$  & $
\sum\limits_{s'=0}^s \frac{\alpha_z^{s-s'} \alpha^{i_1} \ldots
\alpha^{i_{s'}}}{s'!\sqrt{(s - s')!}} \, \phi^{i_1\ldots i_{s'}} |0\rangle$, \hspace{0.5cm} $(N_\alpha + N_z-s)\phik=0$
\\[2mm]
field in $AdS_{d+1} $ & {\small $ s'=0,1,\ldots, s $} &
\\ [1mm]\hline
&&
\\[-3mm]
massive spin-$s$ & $\phi_\lambda^{i_1\ldots i_{s'}}$  & $
\sum\limits_{s'=0}^s\,\,\sum\limits_{\lambda\in [s-s']_2}
\frac{\zeta_{\phantom{z}}^{\frac{s-s'+ \lambda}{2}}
\alpha_z^{\frac{s-s'- \lambda}{2}}\alpha^{i_1}\ldots
\alpha^{i_{s'}}}{s'!\sqrt{(\frac{s-s'+ \lambda}{2})! (\frac{s-s'- \lambda}{2})!}}
\, \phi_\lambda^{i_1\ldots i_{s'}} |0\rangle$
\\[2mm]
field in $AdS_{d+1} $ & {\small $ s'=0,1,\ldots, s $} &
\\ [1mm]
& {\small $\lambda \in [s-s']_2$ }&  \hspace{2cm} $(N_\alpha + N_z+ N_\zeta-s)\phik=0$
\\ [1mm]\hline
&&
\\[-3mm]
conformal spin-$s$ & $\phi_{k'}^{i_1\ldots i_{s'}}$  & $
\sum\limits_{s'=0}^s \sum\limits_{k'\in
[k_{s'}]_2} \!\!\! \frac{\zeta^{s-s'}
(\upsilon^\ominussm)^{^{\frac{k_{s'}+k'}{2}}}
(\upsilon^\oplussm)^{^{\frac{k_{s'} - k'}{2}}}}{s'!
\sqrt{(s-s')!}(\frac{k_{s'} + k'}{2})!}\alpha^{i_1} \ldots
\alpha^{i_{s'}} \, \phi_{k'}^{i_1\ldots i_{s'}}|0\rangle$
\\[2mm]
field in $R^{d-1,1}$ &  {\small $ s'=0,1,\ldots, s $}  &
\\ [1mm]
& {\small $ \hspace{-0.8cm}  k' \in [k_{s'}]_2$ }  &  \hspace{0.3cm} $(N_\alpha + N_\zeta-s)\phik=0$,  \hspace{0.5cm} $(N_\zeta + N_\upsilon  - k_s)\phik=0$
\\ [1mm]
& {\small $k_{s'} \equiv s' + \frac{d-6}{2}$ } &
\\ [1mm]\hline
&&
\\[-3mm]
canonical spin-$s$ & $\phi_\cur^{i_1\ldots i_{s'}}$  & $
\sum\limits_{s'=0}^s \frac{\alpha_z^{s-s'} \alpha^{i_1} \ldots
\alpha^{i_{s'}}}{s'!\sqrt{(s - s')!}} \, \phi_\cur^{i_1\ldots i_{s'}} |0\rangle$,  \hspace{0.5cm} $(N_\alpha + N_z - s)\phik=0$
\\[2mm]
current in $R^{d-1,1}$& {\small $ s'=0,1,\ldots, s $} &
\\ [1mm]\hline
&&
\\[-3mm]
canonical spin-$s$ & $\phi_\sh^{i_1\ldots i_{s'}}$  & $
\sum\limits_{s'=0}^s \frac{\alpha_z^{s-s'} \alpha^{i_1} \ldots
\alpha^{i_{s'}}}{s'!\sqrt{(s - s')!}} \, \phi_\sh^{i_1\ldots i_{s'}} |0\rangle$,  \hspace{0.5cm} $(N_\alpha + N_z - s)\phik=0$
\\[2mm]
shadow in $R^{d-1,1}$ & {\small $ s'=0,1,\ldots, s $} &
\\ [1mm]\hline
&&
\\[-3mm]
anomalous spin-$s$ & $\phi_{\cur, \lambda}^{i_1\ldots i_{s'}}$  & $
\sum\limits_{s'=0}^s\,\,\sum\limits_{\lambda \in [s-s']_2}
\frac{\zeta_{\phantom{z}}^{\frac{s-s'+\lambda}{2}}
\alpha_z^{\frac{s-s'-\lambda}{2}}\alpha^{i_1}\ldots
\alpha^{i_{s'}}}{s'!\sqrt{(\frac{s-s'+\lambda}{2})! (\frac{s-s'-\lambda}{2})!}}
\, \phi_{\cur, \lambda}^{i_1\ldots i_{s'}} |0\rangle$
\\[2mm]
current in $R^{d-1,1}$& {\small $ s'=0,1,\ldots, s $} &
\\ [1mm]
& {\small $\lambda \in [s-s']_2$ }& \hspace{2cm} $(N_\alpha + N_z+ N_\zeta-s)\phik=0$
\\ [1mm]\hline
&&
\\[-3mm]
anomalous spin-$s$ & $\phi_{\sh, \lambda}^{i_1\ldots i_{s'}}$  & $
\sum\limits_{s'=0}^s\,\,\sum\limits_{\lambda \in [s-s']_2}
\frac{\zeta_{\phantom{z}}^{\frac{s-s'-\lambda}{2}}
\alpha_z^{\frac{s-s'+\lambda}{2}}\alpha^{i_1}\ldots
\alpha^{i_{s'}}}{s'!\sqrt{(\frac{s-s'+\lambda}{2})! (\frac{s-s'-\lambda}{2})!}}
\, \phi_{\sh, \lambda}^{i_1\ldots i_{s'}} |0\rangle$
\\[2mm]
shadow in $R^{d-1,1}$& {\small $ s'=0,1,\ldots, s $} &
\\ [1mm]
& {\small $\lambda \in [s-s']_2$ }& \hspace{2cm} $(N_\alpha + N_z+ N_\zeta-s)\phik=0$
\\ [1mm]\hline
\end{tabular}
\end{center}

\newpage

{\bf Light-cone gauge action of fields}. Light-cone gauge action and Lagrangian for fields in flat and AdS spaces we found take the form
\beq
&& S = \int dx^+\, d^{d-1} x \, \LL \,,
\hspace{3.4cm} \hbox{ for fields in} \ R^{d-1,1}\,,
\nonumber\\[-10pt]
\label{25112011-44} &&
\\[-10pt]
 && S = \int dx^+\,d^{d-1} x dz  \, \LL \,,
\hspace{3cm} \hbox{ for fields in } \ AdS_{d+1}\,,
\nonumber
\eeq
$d^{d-1} x \equiv dx^- d^{d-2} x$, where the coordinate $x^+$ is considered as an evolution parameter and light-cone gauge Lagrangian $\LL$ is given by
\be \label{25112011-45}
\LL =  \half \phibr (\Box - \MM^2)\phik\,, \qquad \Box = 2\partial^+\partial^- + \partial^i\partial^i\,.
\ee
Operator $\MM^2$ in \rf{25112011-45} takes the same form as the one entering the gauge invariant approach in Table II. From \rf{25112011-45} and Table II, we see that, for the case of AdS fields, our light-cone gauge Lagrangian \rf{25112011-45} leads to decoupled equations of motion  which  are easily solved in terms of the Bessel function. Our light-cone gauge notation may be found in Appendix.%
\footnote{ Methods for solving equations of motion for higher-spin fields without gauge fixing are discussed in Refs.\cite{Didenko:2009td}.}

Lagrangian \rf{25112011-45} implies the standard equal-time Poisson-Dirac brackets,
\beq
&& \hspace{-0.7cm} [|\phi(x)\rangle,\langle\phi(x')|]\Bigr|_{{\rm equal} \ x^+} =
-\frac{1}{2\partial^+} \delta^{d-1}(x-x')|\rangle \langle |  \,, \hspace{3cm} \hbox{ for fields in } \ R^{d-1,1}\,,
\nonumber\\[-10pt]
\label{mna02-14122013-01} &&
\\[-10pt]
&& \hspace{-0.7cm} [|\phi(x,z)\rangle,\langle\phi(x',z')|]\Bigr|_{{\rm equal} \ x^+} = -\frac{1}{2\partial^+}
\delta^{d-1}(x-x')\delta(z-z') |\rangle \langle |  \,, \hspace{0.5cm} \hbox{ for fields
in } \ AdS_{d+1}\,,\qquad
\nonumber
\eeq
where we show explicit dependence of the ket-vector on the space-time coordinates and the notation $|\rangle\langle|$ stands for the corresponding unit operator on space of the traceless ket-vectors \rf{man02-18122013-01}.

To summarize, starting with the gauge invariant Lagrangian for the double-traceless arbitrary spin fields \rf{man02-26112011-11xx} we obtained the light-cone gauge action in terms of fields which are traceless tensor fields of the $so(d-2)$ algebra.

\noindent {\bf Light-cone gauge 2-point vertices of currents and shadows}. In light-cone gauge, the gauge invariant 2-point vertices given in \rf{man02-15122013-07}, \rf{man02-15122013-08} take the form
\beq
\label{24052014-03} && \Gamma^{\rm cur-\sh} = \int d^d x\,  \LL^{\rm cur-sh}\,, \qquad \LL^{\rm cur-sh} = \langle \phi_\cur(x)| \phi_\sh(x)\rangle \,,
\\
\label{24052014-04} && \Gamma^{\rm sh-sh}= \int d^dx_1 d^dx_2\, \LL_{12}^{\rm sh-sh} \,,
\\
&&  \label{24052014-04x1} \hspace{1.8cm} \LL_{12}^{\rm sh-sh} \equiv \half \langle\phi_\sh(x_1)|
\frac{  f_\nu}{ |x_{12}|^{2\nu + d }} |\phi_\sh (x_2)\rangle \,,
\\
&&  \hspace{1.8cm} |x_{12}|^2 \equiv 2x_{12}^+ x_{12}^- + x_{12}^i x_{12}^i\,, \qquad x_{12}^a = x_1^a - x_2^a\,,
\eeq
where the operators $f_\nu$, $\nu$ take the same form as in \rf{man02-15122013-09}, \rf{man02-15122013-10}, \rf{man02-15122013-11}.

{\bf Massless arbitrary spin field in $AdS_4$}. We now discuss some simplification of Lagrangian \rf{25112011-45} for the case massless field in $AdS_4$. For the case of $AdS_4$, we have $d=3$ and therefore the vector index of $so(d-2)$ algebra $i$ take only one value, $i=1$. Therefore the traceless constraint \rf{man02-18122013-01} implies that $\phik$ can be presented as
\beq
\label{21052014-01} && \phik = |\phi_s\rangle + \alpha^1 |\phi_{s-1}\rangle\,,
\\
\label{21052014-02} && N_z |\phi_s\rangle = s |\phi_s\rangle\,,\qquad N_z |\phi_{s-1}\rangle = (s-1) |\phi_{s-1}\rangle\,,
\eeq
where new ket-vectors $|\phi_s\rangle$, $|\phi_{s-1}\rangle$ are independent of the oscillator $\alpha^1$ and depend only on the oscillator $\alpha^z$. Relations \rf{21052014-02} tell us that the ket-vectors $|\phi_s\rangle$  and $|\phi_{s-1}\rangle$ are the respective degree-$s$ and degree-$(s-1)$ homogeneous monomials in the oscillator $\alpha^z$. Note that relations \rf{21052014-02} are obtained from \rf{21052014-01} and the fact that the ket-vector $\phik$ is homogeneous degree-$s$ polynomial in the oscillators $\alpha^1$, $\alpha^z$ (see Table V). In turn, relations \rf{21052014-02} and expressions for the operator $\MM^2$ for massless fields in $AdS_4$ given in Table II imply that the operator $\MM^2$ is simplified as
\be \label{21052014-03}
- \MM^2 \phik = \partial_z^2\phik\,.
\ee
Using \rf{21052014-03} in \rf{25112011-45}, we get, as in Ref.\cite{Metsaev:1999ui}, the following simple light-cone gauge Lagrangian for massless arbitrary spin field in $AdS_4$,
\be  \label{21052014-04}
\LL =  \half \phibr (\Box + \partial_z^2)\phik\,, \hspace{1cm}  \hbox{ for massless field in } AdS_4\,.
\ee

The following remarks are in order.

\noindent {\bf i}) For the derivation of our light-cone gauge formulation, we use gauge invariant approach discussed in Secs.\ref{section-02},\ref{section-03}. The gauge invariant approach is formulated in terms of Lorentz algebra ket-vector $|\phibf\rangle$. To develop light-cone gauge formulation of fields, currents, and shadows  we impose the following standard light-cone gauge condition on the ket-vector $|\phibf\rangle$:
\be \label{20052014-01}
\bar\alpha^+ \Pibf^\smponetwo |\phibf\rangle = 0\,,
\ee
where $\Pibf^\smponetwo$ is given in \rf{man02-17122013-02a}. Using \rf{20052014-01} and constraints obtained from action \rf{25112011-44}, we get the following relations for gauge-fixed ket-vector $|\phibf\rangle$ and light-cone ket-vector $\phik$,
\beq
&& \label{01062014-01} |\phibf\rangle = \exp\bigl(-\frac{\alpha^+}{\partial^+}\bar\alpha^i\partial^i +
\frac{\alpha^+}{\partial^+} \eb_1 \bigr) |\phi\rangle\,, \qquad  \phik = |\phibf\rangle\bigr|_{\alpha^+=0,\alpha^-=0}\,.
\eeq
Note that, for the case of currents and shadows, in order to get relations in \rf{01062014-01} we should use differential constraint \rf{man02-12122013-03}. Method for the derivation of the light-cone gauge formulation from the gauge invariant formulation is the same as the one discussed in Sec.3.6 in Ref.\cite{Metsaev:1999ui}. Note that, in Ref.\cite{Metsaev:1999ui}, we dealt with massless fields in AdS space. Remarkable feature of the gauge invariant approach we reviewed in Secs.\ref{section-02},\ref{section-03} in this paper is that our approach allows us to treat all fields (massless, massive, and conformal) in flat and AdS spaces on an equal footing. This is the reason why the method in Ref.\cite{Metsaev:1999ui} is generalized to all the fields considered in this paper in a rather straightforward way.

\noindent {\bf ii}) For massless fields in $AdS_{d+1}$, $d>3$, and massive fields in $AdS_{d+1}$, $d\geq 3$, results in this paper provide an alternative light-cone gauge formulation as compared the one in Refs.\cite{Metsaev:1999ui,Metsaev:2003cu}. Advantage of light-cone formulation of AdS field dynamics we develop in this paper as compared the one in Refs.\cite{Metsaev:1999ui,Metsaev:2003cu} is that light-cone gauge action obtained in this paper leads to decoupled equations of motion which are easily solved in terms of Bessel functions. Light-cone gauge formulation of conformal fields dynamics developed in this paper has not been discussed in the earlier literature.

\noindent {\bf iii)} In the framework of AdS/CFT correspondence, Euclidean signature light-cone gauge Lagrangian, which involves a proper boundary term, and solution to Dirichlet problem for equations of motion of the light-cone gauge AdS field $\phik$ with boundary conditions corresponding to the light-cone gauge boundary shadow $|\phi_\sh\rangle$ can be presented as
\beq
 \label{01062014-01z1}  \LL^{E}  & = &   \half \langle \partial^a \phi| | \partial^a \phi\rangle +\half
\langle \TT_{\nu-\half} \phi|     | \TT_{\nu-\half} \phi\rangle\,,
\\
\label{01062014-01z2} |\phi(x,z)\rangle  & = &  \sigma_\nu \int d^dy\, G_\nu (x-y,z)
|\phi_\sh(y)\rangle\,,
\\
&& G_\nu(x,z) = \frac{c_\nu z^{\nu+\half}}{ (z^2+
|x|^2)^{\nu + \frac{d}{2}} }\,,   \quad  c_\nu \equiv \frac{\Gamma(\nu+\frac{d}{2})}{\pi^{d/2} \Gamma(\nu)} \,,\quad
\sigma_\nu \equiv \frac{2^\nu\Gamma(\nu)}{
2^{\bar\kappa}\Gamma(\bar\kappa)}(-)^{N_z} \,.\qquad
\eeq
In \rf{01062014-01z2}, boundary canonical shadow enters the solution of the Dirichlet problem for massless AdS field, while the boundary  anomalous shadow enters the solution of the Dirichlet problem for massive AdS field. The $\nu$ and $\bar\kappa$ are defined in \rf{man02-15122013-10},\rf{man02-15122013-11}. Plugging \rf{01062014-01z1}, \rf{01062014-01z2} in \rf{25112011-44}, gives action of AdS fields evaluated on the solution of the Dirichlet problem,
\beq
-S_\eff  =  2{\bar\kappa} c_{\bar\kappa} \Gamma^{\sh-\sh} \,,
\eeq
where the $\Gamma^{\sh-\sh}$ is defined in \rf{24052014-04},\rf{24052014-04x1}.

\newsection{ \large Light-cone gauge realization of relativistic symmetries } \label{section-06}

Algebras of relativistic symmetries of fields, currents, and shadows  in $R^{d-1,1}$
and fields in $AdS_{d+1}$ contain the Lorentz subalgebra $so(d-1,1)$.
In light-cone approach, the Lorentz symmetries $so(d-1,1)$ are not realized manifestly. Therefore, in the framework of the light-cone approach, complete description of field dynamics implies that we have to work out explicit realization of the Lorentz symmetries
$so(d-1,1)$ and the remaining relativistic symmetries as well. We now
discuss the realization of relativistic symmetries in the framework of light-cone gauge
approach.

In light-cone approach, the Poincar\'e algebra  generators can be
separated into two groups:
\beq
&& P^i,\ \ P^+, \ \ J^{+i}, \ \ J^{+-}, \ \ J^{ij}, \hspace{2cm} \hbox{ kinematical generators};
\\
&& P^-,\ \ J^{-i}\,, \hspace{4.8cm} \hbox{ dynamical
generators}.
\eeq
In order to discuss relativistic symmetries of AdS and conformal fields, the Poincar\'e symmetries should be supplemented by the dilatation symmetry and conformal boost symmetries which also can be separated into two groups,%
\footnote{ Note that, for AdS fields, generators in \rf{man02-17122013-22}, \rf{man02-17122013-23} realize isometry symmetries of $AdS_{d+1}$, while, for conformal fields, these generators realize conformal symmetries of $R^{d-1,1}$.}
\beq
\label{man02-17122013-22} && D, \ \ K^i,\ \ K^+,  \hspace{3cm} \hbox{ kinematical generators};
\\
\label{man02-17122013-23} && K^-\,, \hspace{4.5cm} \hbox{ dynamical generators}.
\eeq
We now consider light-cone gauge realization of relativistic symmetries for fields, currents, and shadows in turn.

\subsection{ Light-cone gauge realization of relativistic symmetries for fields}

Field theoretical representation of relativistic symmetry generators $G_\field$ takes the form
\beq
&& G_\field = \int dx^-d^{d-2}x\,\langle\partial^+\phi|G_\diff|\phi\rangle\,,
\hspace{3cm} \hbox{ for fields in} \ R^{d-1,1}\,,
\nonumber\\[-10pt]
\label{man02-14122013-04} &&
\\[-10pt]
&& G_\field = \int dz dx^-d^{d-2}x\,\langle\partial^+\phi|G_\diff|\phi\rangle,
\hspace{2.6cm} \hbox{ for fields in } \ AdS_{d+1}\,,\qquad
\nonumber
\eeq
where $G_\diff$ stands for the realization of the generators in terms of differential operators acting on the $\phik$. We now present the realization of the generators in terms of the differential operators $G_\diff$ acting on the $\phik$.

\noindent {\bf Kinematical generators},
\beq
\label{man02-17122013-04} && P^i=\partial^i\,, \qquad  P^+=\partial^+\,,
\\
\label{man02-17122013-06} && J^{+-} = x^+P^-  - x^-\partial^+\,,
\\
\label{man02-17122013-07} && J^{+i}=  x^+\partial^i - x^i\partial^+\,,
\\
\label{man02-17122013-08} && J^{ij} = x^i\partial^j-x^j\partial^i + M^{ij}\,,
\\
\label{man02-17122013-09} && D = x^+P^- + x^-\partial^+ + x^i\partial^i+ \Delta\,,
\\
\label{man02-17122013-10} && K^+ = K_\Delta^+  +  r_{1,1}\partial^+\,,
\\
\label{man02-17122013-11} && K^i = K_\Delta^i  + r_{1,1}\partial^i + M^{ij} x^j + M^{i-}x^+ + M^{\ominussm i} \,,
\\
\label{man02-17122013-12} && \hspace{1cm} K_\Delta^a \equiv  - \half (2x^+x^- + x^ix^i)\partial^a + x^a D\,,\qquad a=+,i\,,
\eeq

\noindent {\bf Dynamical generators},
\beq
\label{man02-17122013-16}&& P^- = \frac{-\partial^i\partial^i + \MM^2 }{2\partial^+}\,,
\\
\label{man02-17122013-17} && J^{-i} = x^-\partial^i-x^i P^- + M^{-i}\,,
\\
\label{man02-17122013-18} && K^- = K_\Delta^-  + r_{1,1}P^- + x^i M^{-i} - M^{\ominussm i} \frac{\partial^i}{\partial^+} +  \frac{1}{\partial^+}B   \,,
\\
\label{man02-17122013-19} && \hspace{1cm} K_\Delta^-\equiv  - \half (2x^+x^- + x^ix^i)P^- + x^- D\,,
\\
\label{man02-17122013-20} && \hspace{1cm}  M^{-i} \equiv M^{ij}\frac{\partial^j}{\partial^+}+\frac{1}{\partial^+} M^i \,, \qquad M^{i-} = - M^{-i}\,,
\\
\label{man02-17122013-21} && \hspace{1cm}  M^i \equiv e_1 \bar\alpha^i - A^i \eb_1\,, \hspace{2cm}   [M^i,M^j] = \MM^2 M^{ij}\,,
\\
\label{man02-15122013-02} && \hspace{1cm} M^{\ominussm i} \equiv r_{0,1} \bar\alpha^i + A^i \rb_{0,1}\,, \hspace{1.3cm} B \equiv e_1 \rb_{0,1} + r_{0,1} \eb_1 - N_\alpha\,,
\eeq
where operators $A^i$, $M^{ij}$, $N_\alpha$ are given in \rf{25052014-01}. Operators $\MM^2$, $e_1$, $\eb_1$, $r_{1,1}$, $r_{0,1}$, $\rb_{0,1}$ appearing in  \rf{man02-17122013-09}-\rf{man02-17122013-18} take the same form as the ones in the gauge invariant approach for fields we discussed in Secs.\ref{section-02},\ref{section-04} (see Tables II, IV). The following remarks are in order.

\noindent {\bf i}) Appearance of the ladder operators $\MM^2$, $e_1$, $\eb_1$, $r_{1,1}$, $r_{0,1}$, $\rb_{0,1}$ in light-cone gauge operators \rf{man02-17122013-16}-\rf{man02-15122013-02}, in gauge invariant Lagrangian \rf{man02-26112011-11xx}, and in gauge transformations \rf{man02-26112011-23} demonstrates explicitly how the ladder operators entering the gauge invariant formulation of fields manifest themselves in the light-cone gauge formulation for those fields. We note that it is the use of the ladder operators that allows us to treat light-cone gauge fields in AdS space and conformal fields in flat space on an equal footing.
\\
{\bf ii}) Taking into account equal-time commutation relations given in \rf{mna02-14122013-01}, we make sure that equal time commutator of field with the field theoretical generators \rf{man02-14122013-04} takes the form
\be \label{man02-14122013-02}
[|\phi\rangle,G_\field] = G_\diff|\phi\rangle,
\ee
as it should be. Also we check that action \rf{25112011-44} is invariant under   transformation \rf{man02-14122013-02}.
\\
{\bf iii}) Using the operators $e_1$, $\eb_1$ and $r_{0,1}$, $\rb_{0,1}$  given in Tables II and IV, we get the following alternative form for the operator $B$ \rf{man02-15122013-02}:
\beq
&& B = - s - N_z (2s+d-4-N_z), \hspace{1.4cm} \hbox{for massless spin-$s$ field in $AdS_{d+1}$},
\\
&& B  = - s + N_\zeta(N_\zeta+ \kappa -s  - \frac{d- 4}{2})
\nonumber\\
&& \hspace{1.5cm} + \, N_z(N_z -\kappa -s - \frac{d- 4}{2}), \hspace{0.8cm} \hbox{for massive spin-$s$ field in $AdS_{d+1}$},
\\
&& B  = - s - N_\zeta (2s+d-4-N_\zeta)\,,  \hspace{1.3cm} \hbox{for conformal spin-$s$ field in $R^{d-1,1}$}.\qquad
\eeq

\noindent {\bf iv}) Using generators \rf{man02-17122013-04}-\rf{man02-15122013-02}, we compute the 2nd order Casimir operator for conformal and AdS fields. Our expressions for eigenvalues of the Casimir operator in Table VII coincide with the ones given in textbook. This provides additional check for generators given in \rf{man02-17122013-04}-\rf{man02-15122013-02}.

\subsection{ Light-cone gauge realization of relativistic symmetries for currents and shadows}

Representation for generators of $so(d,2)$ algebra in terms of differential operators acting on light-cone ket-vectors of currents and shadows in Table V is given by
\beq
\label{24052014-01} && P^+=\partial^+\,,\quad P^i = \partial^i\,,\quad P^- = \partial^-\,,
\\
&& J^{+-} = x^+ \partial^-  - x^-\partial^+\,,
\\
&& J^{+i}=  x^+\partial^i - x^i\partial^+\,,
\\
&& J^{ij} = x^i\partial^j-x^j\partial^i + M^{ij}\,,
\\
&& J^{-i} = x^-\partial^i - x^i\partial^- + M^{-i}\,,
\\
\label{man02-15122013-03} && D = x^+ \partial^- + x^-\partial^+ + x^i\partial^i+ \Delta\,,
\\
&& K^+ = K_\Delta^+\,,
\\
\label{man02-15122013-04}  && K^i = K_\Delta^i + M^{ij} x^j + \half\{ M^{i-},x^+\} + M^{\ominussm i}\,,
\\
\label{man02-15122013-05}  && K^- = K_\Delta^- + \half \{M^{-i}, x^i\}  - M^{\ominussm i}\frac{\partial^i}{\partial^+} + \frac{1}{\partial^+}B\,,
\\
&& \hspace{1cm}  K_\Delta^a \equiv - \half (2x^+x^-+x^jx^j) \partial^a + x^a D\,,\qquad a=+,-,i\,,
\\
&& \hspace{1cm} M^{-i} \equiv M^{ij} \frac{\partial^j}{\partial^+} + \frac{1}{\partial^+} M^i\,, \qquad M^{i-} = -M^{-i}\,,
\\
\label{man02-15122013-06} && \hspace{1cm} M^i \equiv e_1 \bar\alpha^i - A^i \eb_1\,,
\qquad [M^i,M^j] = \Box M^{ij}\,,
\eeq
where the operators $\Delta$ and $e_1$, $\eb_1$ appearing in \rf{man02-15122013-03} and  \rf{man02-15122013-06} respectively coincide with the ones appearing in gauge invariant formulation of currents and shadows in Tables II, IV. Expressions for operators $M^{\ominussm i}$ and $B$ appearing in \rf{man02-15122013-04}, \rf{man02-15122013-05} are summarized in Table VI. The following remarks are in order.

\noindent {\bf i}) Using notation for $G_\diff$ for differential operators given in \rf{24052014-01}-\rf{man02-15122013-05} and introducing transformation rules for currents and shadows
\be \label{24052014-02}
\delta \phik = G_\diff \phik\,,
\ee
we note that light-cone gauge vertices \rf{24052014-03}, \rf{24052014-04} are invariant under transformation \rf{24052014-02}.

\noindent {\bf ii}) Generators in \rf{24052014-01}-\rf{man02-15122013-05} are polynomial with respect to derivatives $\partial^i$, $\partial^-$. This property of  generators in \rf{24052014-01}-\rf{man02-15122013-05} is the main advantage of light-cone gauge approach to currents and shadows in this paper against the one developed in Refs.\cite{Metsaev:1999ui,Metsaev:2005ws}.

\noindent {\bf iii}) Appearance of the ladder operators $e_1$, $\eb_1$ in light-cone operators \rf{man02-15122013-06} and in covariant differential constraint in \rf{man02-12122013-03} demonstrates explicitly how the ladder operators entering the gauge invariant formulation of currents and shadows manifest themselves in the light-cone gauge formulation for  currents, and shadows.

\noindent {\bf iv}) Using generators \rf{24052014-01}-\rf{man02-15122013-06}, we compute the 2nd order Casimir operator for currents and shadows. Our expressions for eigenvalues of the Casimir operator in Table VII coincide with the ones given in textbook. This provides additional check for generators given in \rf{24052014-01}-\rf{man02-15122013-06}.

\bigskip
\bigskip

{\small
\noindent {\sf Table VI. Operators $M^{\ominussm i}$ and $B$ entering
light-cone gauge $so(d,2)$ algebra generators for currents and shadows in \rf{man02-15122013-04},\rf{man02-15122013-05}. Operators $e_z$, $r_z$, $r_\zeta$, and parameter $\kappa$ are given in Table II.}
\begin{center}
\begin{tabular}{|l|c|l|}
\hline &&
\\[-3mm]
Fields   & Operator $M^{\ominussm i}$   & Operator $B$
\\ [1mm]\hline
&&
\\[-1mm]
canonical spin-$s$ & $- (2\nu- 1) A^i e_z \bar\alpha^z $,  & $ B = e_1 \rb  - N_\alpha $
\\[1mm]
current in $R^{d-1,1}$ &  $\nu\equiv s  + \frac{d-4}{2} - N_z$ & $ \ \ \ \ = - s - N_z (2s+d-4-N_z) $
\\ [1mm]\hline
&&
\\[-1mm]
canonical spin-$s$ & $ \alpha^z e_z (2\nu - 1) \bar\alpha^i$,  & $ B = r \eb_1  - N_\alpha $
\\[1mm]
shadow in $R^{d-1,1}$ &  $\nu\equiv s  + \frac{d-4}{2} - N_z$ & $ \ \ \ \ =  - s - N_z (2s+d-4-N_z)$
\\ [1mm]\hline
&&
\\[-1mm]
anomalous spin-$s$ & $ - \zeta r_\zeta (2\nu + 1) \bar\alpha^i - (2\nu- 1) A^i r_z\bar\alpha^z $,  & $B =  -  2\zeta r_\zeta \nu \eb_1  - 2 e_1 \nu r_z \bar\alpha^z - N_\alpha $
\\[1mm]
current in $R^{d-1,1}$ & $\nu\equiv \kappa + N_\zeta - N_z$ & $ \ \ \ \ \ = - s + N_\zeta(N_\zeta+ \kappa -s  - \frac{d- 4}{2})  $
\\[0mm]
&& $ \ \ \ \ \ + \, N_z(N_z -\kappa -s - \frac{d- 4}{2})$
\\ [1mm]\hline
&&
\\[-1mm]
anomalous spin-$s$ & $  \alpha^z r_z (2\nu - 1) \bar\alpha^i + (2\nu + 1) A^i r_\zeta\bar\zeta $,  & $B =   2\alpha^z r_z \nu \eb_1  + 2 e_1 \nu r_\zeta \bar\zeta - N_\alpha $
\\[1mm]
shadow in $R^{d-1,1}$ & $\nu\equiv \kappa + N_\zeta - N_z$ & $ \ \ \ \ \ = - s + N_\zeta(N_\zeta+ \kappa -s  - \frac{d- 4}{2})  $
\\[2mm]
& & $ \ \ \ \ \  +\, N_z(N_z -\kappa -s - \frac{d- 4}{2})  $
\\ [1mm]\hline
\end{tabular}
\end{center}
}

\newpage
{\small
\noindent {\sf Table VII. Second order $so(d,2)$ algebra Casimir operator $\CC_2$ and its eigenvalues $\langle \CC_2\rangle $ for light-cone gauge fields, currents, and shadows. The parameter $\kappa$ is given in Table II.}
\begin{center}
\begin{tabular}{|l|c|l|}
\hline &&
\\[-3mm]
Fields   & Casimir operator $\CC_2$   & Eigenvalues of $\CC_2$
\\ [1mm]\hline
&&
\\[-3mm]
massless spin-$s$ &   & $  2(s-1)(s+d-2) $
\\[0mm]
field in $AdS_{d+1}$ &   $ \hspace{-0.6cm} \CC_2   \equiv   D(D-d)  -\half J^{ab}J^{ab} - 2K^a P^a $ &
\\ [1mm] \cline{1-1}   \\[-0.5cm]  \cline{3-3}
&&
\\[-3mm]
massive spin-$s$ & $ \hspace{1.4cm} = \Delta(\Delta -d) - 2 B - \half M^{ij}M^{ij} - 2 r_{1,1} \MM^2 $  & $ \kappa^2 + s (s+d-2) - \frac{d^2}{4} $
\\[0mm]
field in $AdS_{d+1}$ &    &
\\ [1mm]\cline{1-1} \\  [-0.5cm]  \cline{3-3}
&&
\\[-3mm]
conformal spin-$s$ &   & $ 2(s-1)(s+d-2) $
\\[0mm]
field in $R^{d-1,1}$ &   &
\\ [2mm]\hline
&&
\\ [-3mm]
canonical spin-$s$ &   & $  2(s-1)(s+d-2) $
\\[0mm]
current $R^{d-1,1}$ &   &
\\ [1mm] \cline{1-1}   \\[-0.5cm]  \cline{3-3}
&&
\\[-3mm]
canonical spin-$s$ & $  \hspace{-0.6cm} \CC_2   \equiv   D(D-d)  -\half J^{ab}J^{ab} - 2K^a P^a $ &
\\[0mm]
shadow $R^{d-1,1}$ &   & $ 2(s-1)(s+d-2) $ %
\\ [1mm]\cline{1-1} \\  [-0.5cm]  \cline{3-3}
&&
\\[-3mm]
anomalous spin-$s$ &  $ = \Delta(\Delta -d) - 2 B - \half M^{ij}M^{ij} $  & $ \kappa^2 + s (s+d-2) - \frac{d^2}{4} $
\\[0mm]
current $R^{d-1,1}$ &   &
\\ [0mm] \cline{1-1}   \\[-0.5cm]  \cline{3-3}
&&
\\[-3mm]
anomalous spin-$s$ &    & $ \kappa^2 + s (s+d-2) - \frac{d^2}{4} $
\\[0mm]
shadow in $R^{d-1,1}$ &  &
\\ [0mm]\hline
\mc{3}{|l|}{ $ \CC_2 =   D(D-d) + J^{+-}(J^{+-} - d +2)  - 2 J^{+i}J^{-i} - \half J^{ij} J^{ij} -  2(K^+ P^- + K^- P^+ + K^i P^i) \phantom{\int\limits_{}^1}$}
\\
\mc{3}{|l|}{ $ \langle \CC_2 \rangle = \varepsilon_0(\varepsilon_0 - d) + s(s+d-2) $, \quad  }
\\
\mc{3}{|l|}{ $\varepsilon_0 = s+d-2$ for massless spin-$s$ field in $AdS_{d+1}$ and for canonical spin-$s$ current  in $R^{d-1,1}$ }
\\
\mc{3}{|l|}{ $\varepsilon_0 = 2-s$ for conformal spin-$s$ field in $R^{d-1,1}$ and for spin-$s$ shadow in $R^{d-1,1}$ }
\\
\mc{3}{|l|}{ $\varepsilon_0 = \frac{d}{2} + \kappa $ for massive spin-$s$ field in $AdS_{d+1}$ and for anomalous spin-$s$ current in $R^{d-1,1}$ }
\\
\mc{3}{|l|}{ $\varepsilon_0 = \frac{d}{2} - \kappa $ for anomalous spin-$s$ shadow in $R^{d-1,1}$ }
\\ [1mm] \hline
\end{tabular}
\end{center}
}

\newsection{ \large Conclusions }\label{section-07}

In conclusion, let us briefly discuss a number of the potentially interesting
applications and generalizations of our light-cone approach.
One potentially interesting application is related to the problem of interactions vertices in higher-spin field theories. Although many methods for building interaction vertices for higher-spin gauge fields are known in the literature (see e.g.
Refs.\cite{Fradkin:1987ks}-\cite{Joung:2011ww}), building interaction
vertices for concrete models of higher-spin field theories is still
a challenging problem. Light-cone gauge approach provides interesting possibilities for the studying interaction vertices of higher-spin field theories. This is to say that, on the one hand,  some systematic methods for building light-cone gauge interaction vertices were developed in Refs.\cite{Metsaev:2007rn,Metsaev:2005ar,Metsaev:1993ap}.
On the other hand, in this paper, we demonstrated that the use of the ladder operators allows us to treat free AdS fields and conformal fields on an equal footing.
Therefore one can expect that results in this paper and in Refs.\cite{Metsaev:2007rn,Metsaev:2005ar,Metsaev:1993ap} will provide new interesting possibilities for studying interaction vertices of AdS and conformal fields on an equal footing.

In this paper, we considered the light-cone gauge action for the bosonic totally symmetric fields. Generalization of our approach to the case of fermionic fields \cite{Metsaev:2013wza}-\cite{Buchbinder:2007vq} and mixed-symmetry fields \cite{Skvortsov:2008vs}-\cite{Buchbinder:2011xw} could also be of interest. We believe also that the light-cone gauge approach we
discussed in this paper might be useful for the study of light-cone gauge AdS string theory  \cite{Metsaev:2000mv,Metsaev:2000yf}.

\medskip

{\bf Acknowledgments}. This work was supported by the RFBR Grant No.11-02-00814.

\setcounter{section}{0}\setcounter{subsection}{0}
\appendix{ \large Notation }

Throughout the paper the notation $\lambda \in [k]_2$
implies that $\lambda =-k,-k+2,-k+4,\ldots,k-4, k-2,k$:
\be \label{sumnot02}
\lambda \in [k]_2 \quad \Longrightarrow \quad \lambda
=-k,-k+2,-k+4,\ldots,k-4, k-2,k\,.
\ee

{\bf Notation in basis of Lorentz algebra $so(d-1,1)$}. Throughout this paper, $x^a$ denotes coordinates in $d$-dimensional flat space-time $R^{d-1,1}$,
while $\partial_a$ denotes derivatives with respect to $x^a$, $\partial_a
\equiv
\partial /
\partial x^a$. The vector indices of the Lorentz algebra $so(d-1,1)$ take the following values: $a,b,c,e=0,1,\ldots ,d-1$. We use the mostly positive flat metric
tensor $\eta^{ab}$. To simplify our expressions, we drop $\eta_{ab}$ in scalar
products, i.e., we use $X^aY^a \equiv \eta_{ab}X^a Y^b$.

We use the creation operators $\alpha^a$, $\alpha^z$,
$\zeta$, $\upsilon^\oplussm$, $\upsilon^\ominussm$ and the
respective annihilation operators $\bar{\alpha}^a$,
$\bar\alpha^z$, $\bar\zeta$, $\bar\upsilon^\ominussm$,
$\bar\upsilon^\oplussm$. These operators, to be
referred to as oscillators, satisfy the commutation relations%
\footnote{ Applications and extensive study of the oscillator formalism may
be fond in Refs.\cite{Boulanger:2008up}.}
\beq
\label{man02-291102011-01xx}
&&{}\hspace{-1cm}  [\bar\alpha^a,\alpha^b]  =  \eta^{ab}\,, \qquad
[\bar\alpha^z, \alpha^z] = 1\,, \qquad [\bar\zeta, \zeta] = 1\,, \qquad
[\bar\upsilon^\oplussm, \upsilon^\ominussm] = 1 \,,  \qquad
[\bar\upsilon^\ominussm, \upsilon^\oplussm] = 1  \,,  \quad
\\
\label{man02-291102011-02xx} && {}\hspace{-1cm} \bar\alpha^a  |0\rangle
= 0\,,\qquad \quad \ \ \bar\alpha^z | 0 \rangle =  0\,,\qquad\quad \
\bar\zeta|0\rangle  =  0\,, \qquad \ \bar\upsilon^\oplussm|0\rangle =
0\,, \qquad \quad  \bar\upsilon^\ominussm| 0 \rangle  = 0\,.  \quad
\eeq
We use the following hermitian conjugation rules for derivatives and
oscillators:
\be \label{03082011-01xx} \partial^{a\dagger} = - \partial^a\,, \qquad
\alpha^{a\dagger}  =  \bar\alpha^a\,, \qquad  \alpha^{z\dagger} =
\bar\alpha^z,  \qquad  \zeta^\dagger =  \bar\zeta, \qquad
\upsilon^{\oplussm\dagger}  =  \bar\upsilon^\oplussm, \qquad
\upsilon^{\ominussm\dagger}  =  \bar\upsilon^\ominussm\,.
\ee
Throughout this paper we use operators constructed out of the derivatives and oscillators,
\beq
\label{manold-31102011-02xx}&&  \Box  =  \partial^a \partial^a\,, \hspace{1.5cm}
\alphabf \partialbf \equiv \alpha^a \partial^a\,,  \hspace{1.4cm}
\bar\alphabf \partialbf \equiv \bar\alpha^a \partial^a,
\\
&& \alphabf^2  \equiv  \alpha^a\alpha^a,  \hspace{1.4cm}   \bar\alphabf^2 \equiv
\label{manold-31102011-02a1x} \bar\alpha^a \bar\alpha^a,\qquad
\\
\label{manold-31102011-05xx} && N_\alphabf
\equiv \alpha^a  \bar\alpha^a,\hspace{1.3cm} N_z  \equiv  \alpha^z  \bar\alpha^z, \hspace{1.5cm} N_\zeta  \equiv  \zeta  \bar\zeta,
\\
&& N_{\upsilon^\oplussm}\equiv \upsilon^\oplussm \bar\upsilon^\ominussm\,, \hspace{1.1cm} N_{\upsilon^\ominussm}\equiv \upsilon^\ominussm \bar\upsilon^\oplussm\,,\hspace{1cm} N_\upsilon \equiv N_{\upsilon^\oplussm} + N_{\upsilon^\ominussm}\,,
\\
\label{man02-17122013-01} && \widetilde\Abf{}^a \equiv \alpha^a - \alphabf^2 \frac{1}{2N_\alphabf + d - 2}\bar\alpha^a\,,
\\
\label{man02-17122013-02} && \bar\Abf_\perp^a \equiv \bar\alpha^a - \half \alpha^a\bar\alphabf^2\,,
\\
&& \label{man02-17122013-02a} \Pibf^\smponetwo = 1 - \alphabf^2\frac{1}{2(2N_\alphabf+d)}\bar\alphabf^2\,, \qquad
\mubf \equiv 1 - \frac{1}{4}\alphabf^2 \bar\alphabf^2\,.
\eeq

{\bf Notation in basis of $so(d-2)$ algebra}. To discuss light-cone gauge formulation we use basis of $so(d-2)$ algebra and decompose $so(d-1)$ algebra vector $X^a$ as $X^+$, $X^-$, $X^i$, $i=1,\ldots,d-2$, where  $X^\pm \equiv (X^{d-1} \pm X^0)/\sqrt{2}$. This implies the following decomposition of the space-time coordinates, derivatives, oscillators, and scalar product:
\beq
&& \hspace{-1.4cm}  x^a = x^+,x^-,x^i,\quad \partial^a = \partial^+, \partial^-,\partial^i,\quad
\partial^+\equiv \partial/\partial x^-,\quad
\partial^-\equiv \partial/\partial x^+,\quad
\partial^i\equiv \partial/\partial x^i,
\\
&& \hspace{-1.4cm} \alpha^a = \alpha^+, \alpha^-,\alpha^i,\quad  \bar\alpha^a =
\bar\alpha^+,\bar\alpha^-, \bar\alpha^i,\quad [\bar\alpha^+,\alpha^-] = 1, \quad [\bar\alpha^-,\alpha^+] = 1, \quad
[\bar\alpha^i,\alpha^j] =\delta^{ij},
\\
&& \hspace{3cm} X^a Y^a = X^+ Y^- + X^- Y^+ + X^i Y^i\,.
\eeq
Vector indices of the algebra $so(d-2)$ take the values $i,j=1,\ldots ,d-2$.
We use operators constructed out of the spatial derivative and oscillators,
\beq
&& \alpha\partial \equiv \alpha^i\partial^i\,,\qquad\quad \bar\alpha\partial
\equiv \bar\alpha^i\partial^i\,,\qquad \alpha^2 \equiv \alpha^i\alpha^i\,,
\qquad\quad   \bar\alpha^2 \equiv \bar\alpha^i\bar\alpha^i\,,\qquad
\\
\label{25052014-01} &&  M^{ij} = \alpha^i \bar\alpha^j - \alpha^j\bar\alpha^i\,,\qquad A^i \equiv \alpha^i - \alpha^2 \frac{1}{2N_\alpha + d - 2}\bar\alpha^i\,, \qquad N_\alpha\equiv\alpha^i\bar\alpha^i\,.\qquad
\eeq


\small
\begin{thebibliography}{30}

\parskip=-3pt


\bibitem{Dirac:1951zz}
  P.~A.~M.~Dirac,
  Can.\ J.\ Math.\  {\bf 3}, 1 (1951).


\bibitem{Mandelstam:1982cb}
  S.~Mandelstam,
  Nucl.\ Phys.\ B {\bf 213}, 149 (1983).
%
\\
%
  M.~B.~Green and J.~H.~Schwarz,
  Phys.\ Lett.\ B {\bf 122}, 143 (1983).
%
\\
%
  L.~Brink, M.~B.~Green and J.~H.~Schwarz,
  Nucl.\ Phys.\ B {\bf 223}, 125 (1983).


\bibitem{kaku}
  M.~Kaku and K.~Kikkawa,
  Phys.\ Rev.\ D {\bf 10}, 1110 (1974).

\bibitem{Green:1983hw}
  M.~B.~Green, J.~H.~Schwarz and L.~Brink,
  Nucl.\ Phys.\ B {\bf 219}, 437 (1983).


\bibitem{Bengtsson:1983pd}
  A.~K.~H.~Bengtsson, I.~Bengtsson and L.~Brink,
  Nucl.\ Phys.\ B {\bf 227}, 31 (1983).
%
\\
%
  A.~K.~H.~Bengtsson, I.~Bengtsson and L.~Brink,
  Nucl.\ Phys.\ B {\bf 227}, 41 (1983).


\bibitem{Metsaev:2007rn}
  R.~R.~Metsaev,
  Nucl.\ Phys.\ B {\bf 859}, 13 (2012)
  [arXiv:0712.3526 [hep-th]].


\bibitem{Metsaev:2005ar}
  R.~R.~Metsaev,
  Nucl.\ Phys.\ B {\bf 759}, 147 (2006)
  [hep-th/0512342].


\bibitem{Brodsky:1997de}
  S.~J.~Brodsky, H.~-C.~Pauli and S.~S.~Pinsky,
  Phys.\ Rept.\  {\bf 301}, 299 (1998)
  [hep-ph/9705477].
%
\\
%
  S.~J.~Brodsky, G.~F.~de Téramond and H.~Gün.~Dosch,
   [arXiv:1310.8648 [hep-ph]].
%
\\
%
  G.~F.~de Teramond and S.~J.~Brodsky,
  [arXiv:1203.4025 [hep-ph]].


\bibitem{Metsaev:1999ui}
  R.~R.~Metsaev,
  Nucl.\ Phys.\ B {\bf 563}, 295 (1999)
  [hep-th/9906217].

\bibitem{Metsaev:2005ws}
  R.~R.~Metsaev,
  Phys.\ Lett.\ B {\bf 636}, 227 (2006)
  [hep-th/0512330].


\bibitem{Metsaev:1999gz}
  R.~R.~Metsaev,
  Phys.\ Lett.\ B {\bf 468}, 65 (1999)
  [hep-th/9908114].



\bibitem{Metsaev:2002vr}
  R.~R.~Metsaev,
  Phys.\ Lett.\ B {\bf 531}, 152 (2002)
  [hep-th/0201226].


\bibitem{Metsaev:2003cu}
  R.~R.~Metsaev,
  Phys. Lett.  B {\bf 590}, 95 (2004)
  hep-th/0312297

\bibitem{Metsaev:2004ee}
  R.~R.~Metsaev,
  Class.\ Quant.\ Grav.\  {\bf 22}, 2777 (2005)
  [arXiv:hep-th/0412311].




\bibitem{Koch:2010cy}
  R.d.M.Koch, A.Jevicki, K.Jin and J.P.Rodrigues,
  Phys.Rev. D {\bf 83}, 025006 (2011)
  [arXiv:1008.0633].
%
\\
%
  R.Koch, A.Jevicki, K.Jin, J.Rodrigues, Q.Ye,
  Class.Quant.Grav.{\bf 30}, 104005 (2013)
  [arXiv:1205.4117].
%
\\
%
  A.~Jevicki, K.~Jin and Q.~Ye,
  J.\ Phys.\ A {\bf 46}, 214005 (2013)
  [arXiv:1212.5215 [hep-th]].



\bibitem{Metsaev:2008ks}
  R.R. Metsaev,
  Phys. \ Lett. \ B {\bf 671}, 128 (2009)
  [arXiv:0808.3945].


\bibitem{Metsaev:2009hp}
  R.R. Metsaev,
  Phys. Lett. \ B \ {\bf 682}, 455 (2010)
  [arXiv:0907.2207].


\bibitem{Metsaev:2007fq}
  R.~R.~Metsaev,
  JHEP \ {\bf 1201}, 064 (2012)
  [arXiv:0707.4437 [hep-th]].

\bibitem{Metsaev:2007rw}
  R.~R.~Metsaev,
  JHEP {\bf 1206}, 062 (2012)
  [arXiv:0709.4392 [hep-th]].



\bibitem{Metsaev:2008fs}
  R.~R.~Metsaev,
  Phys.\ Rev.\  D {\bf 78}, 106010 (2008)
  [arXiv:0805.3472 [hep-th]].




\bibitem{Metsaev:2009ym}
  R.~R.~Metsaev,
  Phys.\ Rev.\  D {\bf 81}, 106002 (2010)
  [arXiv:0907.4678 [hep-th]].



\bibitem{Metsaev:2010zu}
  R.~R.~Metsaev,
  Phys.\ Rev.\  D {\bf 83}, 106004 (2011)
  [arXiv:1011.4261 [hep-th]].

\bibitem{Metsaev:2011uy}
  R.~R.~Metsaev,
  Phys.\ Rev.\ D {\bf 85}, 126011 (2012)
  [arXiv:1110.3749 [hep-th]].



\bibitem{Siegel:1999ew}
  W.~Siegel, ``Fields,''
  hep-th/9912205.
%
\\
%
  G.~Barnich, G.~Bonelli and M.~Grigoriev,
  ``From BRST to light-cone description of higher spin gauge fields,''
  hep-th/0502232.



\bibitem{Fronsdal:1978rb}
  C.~Fronsdal,
  Phys.\ Rev.\  D {\bf 18}, 3624 (1978).




\bibitem{Zinoviev:2001dt}
  Yu.~M.~Zinoviev,
  ``On massive high spin particles in (A)dS,''
  arXiv:hep-th/0108192.



\bibitem{Fronsdal:1978vb}
C.~Fronsdal,
Phys.\ Rev.\ D {\bf 20}, 848 (1979).








\bibitem{Aragone:1988yx}
  C.~Aragone, S.~Deser and Z.~Yang,
  Annals Phys.\  {\bf 179}, 76 (1987).


\bibitem{Rindani:1988gb}
  S.~D.~Rindani, D.~Sahdev and M.~Sivakumar,
  Mod.\ Phys.\ Lett.\  A {\bf 4}, 265 (1989).



\bibitem{Buchbinder:2005ua}
  I.~L.~Buchbinder and V.~A.~Krykhtin,
  Nucl.\ Phys.\ B {\bf 727}, 537 (2005)
  [arXiv:hep-th/0505092].



\bibitem{Lopatin:1987hz}
  V.~E.~Lopatin and M.~A.~Vasiliev,
  Mod.\ Phys.\ Lett.\  A {\bf 3}, 257 (1988).



\bibitem{Buchbinder:2001bs}
  I.~L.~Buchbinder, A.~Pashnev and M.~Tsulaia,
  Phys.\ Lett.\  B {\bf 523}, 338 (2001)
  [arXiv:hep-th/0109067].







\bibitem{Sagnotti:2003qa}
  A.~Sagnotti and M.~Tsulaia,
  Nucl.\ Phys.\  B {\bf 682}, 83 (2004)
  [arXiv:hep-th/0311257].


\bibitem{Francia:2005bu}
  D.~Francia and A.~Sagnotti,
  Phys.\ Lett.\ B {\bf 624}, 93 (2005)
  [hep-th/0507144].


\bibitem{Francia:2007qt}
  D.~Francia, J.~Mourad and A.~Sagnotti,
  Nucl.\ Phys.\ B {\bf 773}, 203 (2007)
  [hep-th/0701163].


\bibitem{Campoleoni:2008jq}
  A.Campoleoni, D.Francia, J.Mourad and A.Sagnotti,
  Nucl.Phys.B {\bf 815}, 289 (2009)
  [arXiv:0810.4350].


\bibitem{Francia:2010qp}
  D.~Francia,
  Phys.\ Lett.\ B {\bf 690}, 90 (2010)
  [arXiv:1001.5003 [hep-th]].
%
\\
%
  D.~Francia,
  J.\ Phys.\ Conf.\ Ser.\  {\bf 222}, 012002 (2010)
  [arXiv:1001.3854 [hep-th]].



\bibitem{Buchbinder:2007ak}
  I.L.~Buchbinder, A.V.~Galajinsky and V.~A.~Krykhtin,
  Nucl.Phys. B{\bf 779}, 155 (2007)
  [hep-th/0702161].

\bibitem{Buchbinder:2008ss}
  I.~L.~Buchbinder and A.~V.~Galajinsky,
  JHEP {\bf 0811}, 081 (2008)
  [arXiv:0810.2852 [hep-th]].





\bibitem{Zinoviev:2008ze}
  Yu.~M.~Zinoviev,
  Nucl.\ Phys.\  B {\bf 808}, 185 (2009)
  [arXiv:0808.1778 [hep-th]].


\bibitem{Ponomarev:2010st}
  D.~S.~Ponomarev and M.~A.~Vasiliev,
  Nucl.\ Phys.\  B {\bf 839}, 466 (2010)
  [arXiv:1001.0062 [hep-th]].




\bibitem{Buchbinder:2006ge}
  I.~L.~Buchbinder, V.~A.~Krykhtin and P.~M.~Lavrov,
  Nucl.\ Phys.\  B {\bf 762}, 344 (2007)
  hep-th/0608005


\bibitem{Alkalaev:2011zv}
  K.~Alkalaev and M.~Grigoriev,
  Nucl.\ Phys.\  B {\bf 853}, 663 (2011)
  [arXiv:1105.6111 [hep-th]].


\bibitem{Grigoriev:2011gp}
  M.~Grigoriev and A.~Waldron,
  Nucl.\ Phys.\  B {\bf 853}, 291 (2011)
  [arXiv:1104.4994 [hep-th]].



\bibitem{Metsaev:2000qb}
  R.~R.~Metsaev,
  Nucl.\ Phys.\ Proc.\ Suppl.\  {\bf 102}, 100 (2001)
  [arXiv:hep-th/0103088].


\bibitem{Biswas:2002nk}
  T.~Biswas and W.~Siegel,
  JHEP {\bf 0207}, 005 (2002)
  [arXiv:hep-th/0203115].


\bibitem{Hallowell:2005np}
  K.~Hallowell and A.~Waldron,
  Nucl.\ Phys.\ B {\bf 724}, 453 (2005)
  [hep-th/0505255].

\bibitem{Artsukevich:2008vy}
  A.~Y.~Artsukevich and M.~A.~Vasiliev,
  Phys.\ Rev.\  D {\bf 79}, 045007 (2009)
  [arXiv:0810.2065 [hep-th]].




\bibitem{Giombi:2013yva}
  S.Giombi, I.Klebanov, S.Pufu, B.Safdi, G.Tarnopolsky,
  JHEP {\bf 1310}, 016 (2013)
  [arXiv:1306.5242].


\bibitem{Tseytlin:2013jya}
  A.~A.~Tseytlin,
  Nucl.\ Phys.\ B {\bf 877}, 598 (2013)
  [arXiv:1309.0785 [hep-th]].
%
\\
%
  A.~A.~Tseytlin,
  Nucl.\ Phys.\ B {\bf 877}, 632 (2013)
  [arXiv:1310.1795 [hep-th]].






\bibitem{Guttenberg:2008qe}
S.~Guttenberg and G.~Savvidy,
SIGMAP bulletin 4, 061 (2008) arXiv:0804.0522 [hep-th].


\bibitem{Fotopoulos:2009iw}
  A.~Fotopoulos and M.~Tsulaia,
  JHEP {\bf 0910}, 050 (2009)
  [arXiv:0907.4061 [hep-th]].



\bibitem{Manvelyan:2008ks}
  R.~Manvelyan, K.~Mkrtchyan and W.~Ruhl,
  Nucl.\ Phys.\  B {\bf 803}, 405 (2008)
  [arXiv:0804.1211 [hep-th]].



\bibitem{Chang:2011mz}
  C.~-M.~Chang and X.~Yin,
  JHEP {\bf 1210}, 024 (2012)
  [arXiv:1106.2580 [hep-th]].


\bibitem{Gover:2008sw}
  A.~R.~Gover, A.~Shaukat and A.~Waldron,
  Nucl.\ Phys.\  B {\bf 812}, 424 (2009)
  [arXiv:0810.2867 [hep-th]].


\bibitem{Gover:2008pt}
  A.~R.~Gover, A.~Shaukat and A.~Waldron,
  Phys.\ Lett.\  B {\bf 675}, 93 (2009)
  [arXiv:0812.3364 [hep-th]].
%
\\
%
  M.~Grigoriev and A.~Waldron,
  Nucl.\ Phys.\ B {\bf 853}, 291 (2011)
  [arXiv:1104.4994 [hep-th]].
%
\\
%
  A.~Rod Gover, E.~Latini and A.~Waldron,
  [arXiv:1205.3489 [math.DG]].
%
\\
  A.~R.~Gover and A.~Waldron,
  [arXiv:1104.2991 [math.DG]].
%
\\
%
  E.~Joung, M.~Taronna and A.~Waldron,
  JHEP {\bf 1307}, 186 (2013)
  [arXiv:1305.5809 [hep-th]].


\bibitem{Shaynkman:2004vu}
  O.V.Shaynkman, I.Y.Tipunin and M.A.Vasiliev,
  Rev.Math.Phys. {\bf 18}, 823 (2006)
  [hep-th/0401086].
%
\\
%
  V.~K.~Dobrev,
  Rev.\ Math.\ Phys.\  {\bf 20}, 407 (2008)
  [hep-th/0702152].
%
\\
%
  V.~K.~Dobrev,
  JHEP {\bf 1302}, 015 (2013)
  [arXiv:1208.0409 [hep-th]].
%
\\
%
  V.~K.~Dobrev,
  arXiv:1210.8067 [math-ph].
%
\\
%
  K.~Alkalaev,
  J.\ Phys.\ A {\bf 46}, 214007 (2013)
  [arXiv:1207.1079 [hep-th]].




\bibitem{Metsaev:1995re}
  R.~R.~Metsaev,
  Phys.\ Lett.\ B {\bf 354}, 78 (1995).


\bibitem{Alkalaev:2009vm}
  K.~B.~Alkalaev and M.~Grigoriev,
  Nucl.\ Phys.\ B {\bf 835}, 197 (2010)
  [arXiv:0910.2690 [hep-th]].

\bibitem{Bekaert:2009fg}
  X.~Bekaert and M.~Grigoriev,
  SIGMA {\bf 6}, 038 (2010)
  [arXiv:0907.3195 [hep-th]].
%
\\
%
  X.~Bekaert and M.~Grigoriev,
  Nucl.\ Phys.\ B {\bf 876}, 667 (2013)
  [arXiv:1305.0162 [hep-th]].
%
\\
%
  X.~Bekaert and M.~Grigoriev,
  J.\ Phys.\ A {\bf 46}, 214008 (2013)
  [arXiv:1207.3439 [hep-th]].


\bibitem{Bonezzi:2010jr}
  R.~Bonezzi, E.~Latini and A.~Waldron,
  Phys.\ Rev.\  D {\bf 82}, 064037 (2010)
  [arXiv:1007.1724 [hep-th]].




\bibitem{Fotopoulos:2006ci}
  A.~Fotopoulos, K.~L.~Panigrahi and M.~Tsulaia,
  Phys.\ Rev.\  D {\bf 74}, 085029 (2006)
  [hep-th/0607248].


\bibitem{Didenko:2009td}
  V.~E.~Didenko and M.~A.~Vasiliev,
  Phys.\ Lett.\  B {\bf 682}, 305 (2009)
  [arXiv:0906.3898 [hep-th]].
%
\\
%
  V.~E.~Didenko,
  Class.\ Quant.\ Grav.\  {\bf 29}, 025009 (2012)
  [arXiv:1108.4321 [hep-th]].



\bibitem{Metsaev:1993ap}
  R.~R.~Metsaev,
  Mod.\ Phys.\ Lett.\  A {\bf 8}, 2413 (1993).



\bibitem{Fradkin:1987ks}
  E.~S.~Fradkin and M.~A.~Vasiliev,
  Phys.\ Lett.\  B {\bf 189}, 89 (1987).




\bibitem{Metsaev:2006ui}
  R.~R.~Metsaev,
  Phys.\ Rev.\  D {\bf 77}, 025032 (2008)
  [arXiv:hep-th/0612279].

\bibitem{Polyakov:2010qs}
  D.~Polyakov,
  Int.\ J.\ Mod.\ Phys.\  A {\bf 25}, 4623 (2010);
  [arXiv:1005.5512 [hep-th]].



\bibitem{Sagnotti:2010at}
  A.~Sagnotti and M.~Taronna,
  Nucl.\ Phys.\  B {\bf 842}, 299 (2011)
  [arXiv:1006.5242 [hep-th]].



\bibitem{Bekaert:2010hp}
  X.~Bekaert, N.~Boulanger and S.~Leclercq,
  J.\ Phys.\ A  {\bf 43}, 185401 (2010)
  [arXiv:1002.0289 [hep-th]].



\bibitem{Manvelyan:2010je}
  R.~Manvelyan, K.~Mkrtchyan and W.~Ruehl,
  Phys.\ Lett.\  B {\bf 696}, 410 (2011)
  [arXiv:1009.1054 [hep-th]].



\bibitem{Boulanger:2011qt}
  N.~Boulanger, E.D.~Skvortsov, Y.M.~Zinoviev,
  J.Phys. A {\bf A44}, 415403 (2011).
  [1107.1872 [hep-th]].



\bibitem{Alkalaev:2010af}
  K.~Alkalaev,
  JHEP {\bf 1103}, 031 (2011)
  [arXiv:1011.6109 [hep-th]].


\bibitem{Boulanger:2011se}
  N.~Boulanger, E.~D.~Skvortsov,
  JHEP {\bf 1109}, 063 (2011).
  [arXiv:1107.5028 [hep-th]].


\bibitem{Vasilev:2011xf}
  M.~A.~Vasiliev,
  Nucl.\ Phys.\ B {\bf 862}, 341 (2012)
  [arXiv:1108.5921 [hep-th]].


\bibitem{Henneaux:2012wg}
  M.~Henneaux, G.~Lucena Gómez and R.~Rahman,
  JHEP {\bf 1208}, 093 (2012)
  [arXiv:1206.1048 [hep-th]].


\bibitem{Joung:2011ww}
  E.~Joung and M.~Taronna,
  Nucl.\ Phys.\ B {\bf 861}, 145 (2012)
  [arXiv:1110.5918 [hep-th]].


\bibitem{Metsaev:2013wza}
  R.~R.~Metsaev,
  [arXiv:1311.7350 [hep-th]].


\bibitem{Metsaev:2006zy}
  R.~R.~Metsaev,
  Phys.\ Lett.\  B {\bf 643}, 205 (2006)
  [arXiv:hep-th/0609029].



\bibitem{Buchbinder:2007vq}
  I.L.Buchbinder, V.Krykhtin and A.Reshetnyak,
  Nucl.\ Phys.\ B {\bf 787}, 211 (2007)
  [hep-th/0703049].



\bibitem{Skvortsov:2008vs}
  E.~D.~Skvortsov,
  JHEP {\bf 0807}, 004 (2008)
  [arXiv:0801.2268 [hep-th]].
%
%
  JHEP {\bf 1001}, 106 (2010)
  [arXiv:0910.3334 [hep-th]].
%
  Nucl.\ Phys.\  B {\bf 808}, 569 (2009)
  [arXiv:0807.0903 [hep-th]].



\bibitem{Zinoviev:2008ve}
  Yu.~M.~Zinoviev,
  Nucl.\ Phys.\  B {\bf 812}, 46 (2009)
  [arXiv:0809.3287 [hep-th]].


\bibitem{Buchbinder:2011xw}
  I.~L.~Buchbinder and A.~Reshetnyak,
  Nucl.\ Phys.\ B {\bf 862}, 270 (2012)
  [arXiv:1110.5044 [hep-th]].
%
\\
%
  A.~Reshetnyak,
  Nucl.\ Phys.\ B {\bf 869}, 523 (2013)
  [arXiv:1211.1273 [hep-th]].
%
\\
%
  P.~Y.~Moshin and A.~A.~Reshetnyak,
  JHEP {\bf 0710}, 040 (2007)
  [arXiv:0707.0386 [hep-th]].




\bibitem{Metsaev:2000mv}
  R.~R.~Metsaev and A.~A.~Tseytlin,
  J.\ Math.\ Phys.\  {\bf 42}, 2987 (2001)
  [arXiv:hep-th/0011191].



\bibitem{Metsaev:2000yf}
  R.~R.~Metsaev and A.~A.~Tseytlin,
  Phys.\ Rev.\  D {\bf 63}, 046002 (2001)
  [arXiv:hep-th/0007036].




\bibitem{Boulanger:2008up}
  N.~Boulanger, C.~Iazeolla and P.~Sundell,
  JHEP {\bf 0907}, 013 (2009)
  [arXiv:0812.3615 [hep-th]].
%
\\
%
  X.~Bekaert and N.~Boulanger,
  Commun.\ Math.\ Phys.\  {\bf 271}, 723 (2007)
  [arXiv:hep-th/0606198].







\end{thebibliography}
\end{document}